\documentclass[prl,reprint,superscriptaddress]{revtex4-1}
\usepackage{color}
\usepackage{graphicx}
\usepackage{amsmath}
\usepackage{array}
\usepackage{hyperref}
\usepackage{float}
\usepackage{lineno}
\usepackage{ulem}
\usepackage{siunitx}
\usepackage[bottom]{footmisc}
\begin{document}

\title{Search for Dark-Matter–Nucleon Interactions with a Dark Mediator in PandaX-4T}


\def\TDLee{Tsung-Dao Lee Institute, Shanghai Jiao Tong University, Shanghai 200240, China}
\def\shKeyLab{School of Physics and Astronomy, Shanghai Jiao Tong University, Key Laboratory for Particle Astrophysics and Cosmology (MoE), Shanghai Key Laboratory for Particle Physics and Cosmology, Shanghai 200240, China}
\def\BUAA{School of Physics, Beihang University, Beijing 102206, China}
\def\BUAALab{Beijing Key Laboratory of Advanced Nuclear Materials and Physics, Beihang University, Beijing 102206, China}
\def\zzu{School of Physics and Microelectronics, Zhengzhou University, Zhengzhou, Henan 450001, China}
\def\USTClab{State Key Laboratory of Particle Detection and Electronics, University of Science and Technology of China, Hefei 230026, China}
\def\USTCdep{Department of Modern Physics, University of Science and Technology of China, Hefei 230026, China}
\def\BUAALab{International Research Center for Nuclei and Particles in the Cosmos and Beijing Key Laboratory of Advanced Nuclear Materials and Physics, Beihang University, Beijing 100191, China}
\def\pku{School of Physics, Peking University, Beijing 100871, China}
\def\YaLongSD{Yalong River Hydropower Development Company, Ltd., 288 Shuanglin Road, Chengdu 610051, China}
\def\IAP{Shanghai Institute of Applied Physics, Chinese Academy of Sciences, Shanghai 201800, China}
\def\CHEPpku{Center for High Energy Physics, Peking University, Beijing 100871, China}
\def\SDUdep{Research Center for Particle Science and Technology, Institute of Frontier and Interdisciplinary Science, Shandong University, Qingdao 266237, Shandong, China}
\def\SDUlab{Key Laboratory of Particle Physics and Particle Irradiation of Ministry of Education, Shandong University, Qingdao 266237, Shandong, China}
\def\SDIAT{Shandong Institute of Advanced Technology, Jinan 250103, Shandong, China}
\def\UMD{Department of Physics, University of Maryland, College Park, Maryland 20742, USA}
\def\MESJTU{School of Mechanical Engineering, Shanghai Jiao Tong University, Shanghai 200240, China}
\def\SYU{School of Physics, Sun Yat-Sen University, Guangzhou 510275, China}
\def\SYUSFI{Sino-French Institute of Nuclear Engineering and Technology, Sun Yat-Sen University, Zhuhai 519082, China}
\def\NKU{School of Physics, Nankai University, Tianjin 300071, China}
\def\YTU{Department of Physics, Yantai University, Yantai 264005, China}
\def\FDU{Key Laboratory of Nuclear Physics and Ion-beam Application (MOE), Institute of Modern Physics, Fudan University, Shanghai 200433, China}
\def\USST{School of Medical Instrument and Food Engineering, University of Shanghai for Science and Technology, Shanghai 200093, China}
\def\SJTUSC{Shanghai Jiao Tong University Sichuan Research Institute, Chengdu 610213, China}
\def\SPEIT{SJTU Paris Elite Institute of Technology, Shanghai Jiao Tong University, Shanghai 200240, China}
\def\NNU{School of Physics and Technology, Nanjing Normal University, Nanjing 210023, China}
\def\SYUzhuhai{School of Physics and Astronomy, Sun Yat-Sen University, Zhuhai 519082, China}
\def\UCR{Department of Physics and Astronomy, University of California, Riverside, California 92507, USA}

\affiliation{\TDLee}
\author{Di Huang}\affiliation{\shKeyLab}
\author{Abdusalam Abdukerim}\affiliation{\shKeyLab}
\author{Zihao Bo}\affiliation{\shKeyLab}
\author{Wei Chen}\affiliation{\shKeyLab}
\author{Xun Chen}\affiliation{\shKeyLab}\affiliation{\SJTUSC}
\author{Chen Cheng}\affiliation{\SYU}
\author{Zhaokan Cheng}\affiliation{\SYUSFI}
\author{Xiangyi Cui}\affiliation{\TDLee}
\author{Yingjie Fan}\affiliation{\YTU}
\author{Deqing Fang}\affiliation{\FDU}
\author{Changbo Fu}\affiliation{\FDU}
\author{Mengting Fu}\affiliation{\pku}
\author{Lisheng Geng}\affiliation{\BUAA}\affiliation{\BUAALab}\affiliation{\zzu}
\author{Karl Giboni}\affiliation{\shKeyLab}
\author{Linhui Gu}\affiliation{\shKeyLab}
\author{Xuyuan Guo}\affiliation{\YaLongSD}
\author{Chencheng Han}\affiliation{\TDLee} 
\author{Ke Han}\affiliation{\shKeyLab}
\author{Changda He}\affiliation{\shKeyLab}
\author{Jinrong He}\affiliation{\YaLongSD}
\author{Yanlin Huang}\affiliation{\USST}
\author{Junting Huang}\affiliation{\shKeyLab}
\author{Zhou Huang}\affiliation{\shKeyLab}
\author{Ruquan Hou}\affiliation{\SJTUSC}
\author{Yu Hou}\affiliation{\MESJTU}
\author{Xiangdong Ji}\affiliation{\UMD}
\author{Yonglin Ju}\affiliation{\MESJTU}
\author{Chenxiang Li}\affiliation{\shKeyLab}
\author{Jiafu Li}\affiliation{\SYU}
\author{Mingchuan Li}\affiliation{\YaLongSD}
\author{Shuaijie Li}\affiliation{\TDLee}
\author{Tao Li}\affiliation{\SYUSFI}
\author{Qing Lin}\affiliation{\USTClab}\affiliation{\USTCdep}
\author{Jianglai Liu}\email[Spokesperson: ]{jianglai.liu@sjtu.edu.cn}\affiliation{\shKeyLab}\affiliation{\TDLee}\affiliation{\SJTUSC}
\author{Congcong Lu}\affiliation{\MESJTU}
\author{Xiaoying Lu}\affiliation{\SDUdep}\affiliation{\SDUlab}
\author{Lingyin Luo}\affiliation{\pku}
\author{Yunyang Luo}\affiliation{\USTCdep}
\author{Wenbo Ma}\affiliation{\shKeyLab}
\author{Yugang Ma}\affiliation{\FDU}
\author{Yajun Mao}\affiliation{\pku}
\author{Yue Meng}\affiliation{\shKeyLab}\affiliation{\SJTUSC}
\author{Xuyang Ning}\affiliation{\shKeyLab}
\author{Ningchun Qi}\affiliation{\YaLongSD}
\author{Zhicheng Qian}\affiliation{\shKeyLab}
\author{Xiangxiang Ren}\affiliation{\SDUdep}\affiliation{\SDUlab}
\author{Nasir Shaheed}\affiliation{\SDUdep}\affiliation{\SDUlab}
\author{Xiaofeng Shang}\affiliation{\shKeyLab}
\author{Xiyuan Shao}\affiliation{\NKU}
\author{Guofang Shen}\affiliation{\BUAA}
\author{Lin Si}\affiliation{\shKeyLab}
\author{Wenliang Sun}\affiliation{\YaLongSD}
\author{Andi Tan}\affiliation{\UMD}
\author{Yi Tao}\affiliation{\shKeyLab}\affiliation{\SJTUSC}
\author{Anqing Wang}\affiliation{\SDUdep}\affiliation{\SDUlab}
\author{Meng Wang}\affiliation{\SDUdep}\affiliation{\SDUlab}
\author{Qiuhong Wang}\affiliation{\FDU}
\author{Shaobo Wang}\affiliation{\shKeyLab}\affiliation{\SPEIT}
\author{Siguang Wang}\affiliation{\pku}
\author{Wei Wang}\affiliation{\SYUSFI}\affiliation{\SYU}
\author{Xiuli Wang}\affiliation{\MESJTU}
\author{Zhou Wang}\affiliation{\shKeyLab}\affiliation{\SJTUSC}\affiliation{\TDLee}
\author{Yuehuan Wei}\affiliation{\SYUSFI}
\author{Mengmeng Wu}\affiliation{\SYU}
\author{Weihao Wu}\affiliation{\shKeyLab}
\author{Jingkai Xia}\affiliation{\shKeyLab}
\author{Mengjiao Xiao}\affiliation{\UMD}
\author{Xiang Xiao}\affiliation{\SYU}
\author{Pengwei Xie}\affiliation{\TDLee}
\author{Binbin Yan}\affiliation{\shKeyLab}
\author{Xiyu Yan}\affiliation{\SYUzhuhai}
\author{Jijun Yang}\affiliation{\shKeyLab}
\author{Yong Yang}\email[Corresponding author: ]{yong.yang@sjtu.edu.cn}\affiliation{\shKeyLab}
\author{Yukun Yao}\affiliation{\shKeyLab}
\author{Chunxu Yu}\affiliation{\NKU}
\author{Ying Yuan}\affiliation{\shKeyLab}
\author{Zhe Yuan}\affiliation{\FDU} %
\author{Xinning Zeng}\affiliation{\shKeyLab}
\author{Dan Zhang}\affiliation{\UMD}
\author{Minzhen Zhang}\affiliation{\shKeyLab}
\author{Peng Zhang}\affiliation{\YaLongSD}
\author{Shibo Zhang}\affiliation{\shKeyLab}
\author{Shu Zhang}\affiliation{\SYU}
\author{Tao Zhang}\affiliation{\shKeyLab}
\author{Wei Zhang}\affiliation{\TDLee}
\author{Yang Zhang}\affiliation{\SDUdep}\affiliation{\SDUlab}
\author{Yingxin Zhang}\affiliation{\SDUdep}\affiliation{\SDUlab} %
\author{Yuanyuan Zhang}\affiliation{\TDLee}
\author{Li Zhao}\affiliation{\shKeyLab}
\author{Qibin Zheng}\affiliation{\USST}
\author{Jifang Zhou}\affiliation{\YaLongSD}
\author{Ning Zhou}\affiliation{\shKeyLab}\affiliation{\SJTUSC}
\author{Xiaopeng Zhou}\affiliation{\BUAA}
\author{Yong Zhou}\affiliation{\YaLongSD}
\author{Yubo Zhou}\affiliation{\shKeyLab}
\collaboration{PandaX Collaboration}
\author{Ran Huo}\email[Corresponding author: ]{huor@iat.cn}\affiliation{\SDIAT}
\author{Haibo Yu}\email[Corresponding author:]{haiboyu@ucr.edu}\affiliation{\UCR}
\noaffiliation

\date{\today}

\begin{abstract}
 We report results of a search for dark-matter-nucleon interactions via a dark mediator using optimized low-energy data from the PandaX-4T liquid xenon experiment. With the ionization-signal-only data and utilizing the Migdal effect, we set the most stringent limits on the cross section for dark matter masses ranging from 30~$\rm{MeV/c^2}$ to 2~$\rm{GeV/c^2}$. Under the assumption that the dark mediator is a dark photon that decays into scalar dark matter pairs in the early Universe, we rule out significant parameter space of such thermal relic dark-matter model.
\end{abstract}

\maketitle
{\it Introduction.}\textemdash
One particularly important question on the dark matter (DM) is how it interacts with standard model (SM) particles beyond the gravitational effect.
From the perspective of particle physics, one popular construction involves a dark force carrier that mediates the interactions between DM and SM particles. In the traditional DM direct search 
experiments~\cite{PandaX-4T:2021bab,LZ:2022lsv,XENON:2023sxq}, a heavy mediator is usually assumed, 
leading to a mediator-mass-independent DM-nucleon cross section.
However, when the mediator mass is comparable to or even smaller than the momentum transfer in nuclear recoils (NRs), a softer recoil spectrum is expected, which leads to mediator-mass-dependent search results~\cite{Ren:2018gyx,PandaX-II:2021lap,PandaX:2023toi,PandaX:2023tfq}. 
The dark mediator can also indirectly interact with SM particles by mixing with gauge or Higgs bosons. If it kinetically mixes with the ordinary photon, it is called a dark photon, which is a well-motivated vector boson arising from a hidden U(1) gauge symmetry. Dark photons can provide possible explanations of anomalies ranging from particle physics to cosmology, e.g., in muon $g-2$~\cite{Muong-2:2006rrc,Muong-2:2021ojo}, the $^8$Be nuclear transitions~\cite{Krasznahorkay:2015iga}, and the so-called small-scale problems in galactic astronomy (see Ref.~\cite{Tulin:2017ara} for an overview). They have been searched extensively at dedicated fixed-target experiments~\cite{Hearty:2022wij,battaglieri2017us,NA64:2023wbi,MiniBooNEDM:2018cxm,LSND:2001akn,Batell:2014mga} and at colliders~\cite{BaBar:2017tiz}. More dedicated experiments are planned~\cite{berlin2019dark,Chen:2022liu}.

In this Letter, we report a highly sensitive search for DM-nucleon interaction mediated by a dark mediator using optimized low-energy data from the commissioning run of the PandaX-4T experiment, and use the results to test a class of dark-photon-mediated DM models. 
PandaX-4T is the third generation DM direct detection experiment of the PandaX project, located at the China Jinping Underground Laboratory. The central apparatus of PandaX-4T is a dual-phase xenon time projection chamber (TPC), which contains 3.7~tonnes of liquid xenon in the sensitive volume. Particle interaction with xenon nuclei or electrons in the liquid xenon produces scintillation photons ($S1$ signal) and the ionized electrons ($S2$ signal). 
Both signals are detected by the top and bottom arrays of 368 Hamamatsu R11410-23 3-inch photomultiplier tubes (PMTs).  
Analog signals from the PMTs are digitized and then read out in a triggerless scheme~\cite{yang2022readout}, using CAEN V1725B digitizers~\footnote{see \url{https://www.caen.it/products/v1725/}.}, which is crucial to enhance the sensitivity to low-energy recoils. 
More detailed descriptions of the PandaX-4T experiment can be found in Ref.~\cite{PandaX-4T:2021bab}.

{\it DM-nucleon interaction with a dark mediator}. \textemdash First, we consider the physics scenario in which a vector or scalar force mediator $\phi$ coherently mediates the interaction between the DM and nucleon, with the same effective couplings for the proton and neutron. The differential NR recoil rate (in the unit of events/day/kg/keV) for elastic scattering between the DM and xenon nucleus is given by~\cite{Savage:2008er,Kaplinghat:2013yxa}
\begin{equation}
\label{eq:drde}
\begin{split}
\frac{dR}{dE_{\rm{NR}}} &= 
\sigma|_{q^2=0}\frac{A^{2}}{\mu_{p}^{2}}\frac{m^{4}_{\phi}}{(m^{2}_{\phi}+q^{2})^2}F^{2}(q^2) \\ &\times
\frac{\rho}{2m_{\chi}}\int_{v\geq v_{\rm{min}}}\frac{f(v)}{v} d^{3}v,
\end{split}
\end{equation}
where $q^2$ is the four-momentum-transfer-squared, $\sigma|_{q^2=0}$ is the DM-nucleon cross section in the limit of zero momentum transfer, $A$ is the xenon mass number, $\mu_p$ is the DM-nucleon reduced mass, $m_\phi (m_\chi)$ is the mediator (DM) mass, $F(q^2)$ is the nuclear form factor, $\rho$ is the local DM density, $f(v)$ is the DM velocity distribution relative to the detector, and $v_{\rm{min}}$ is the minimum DM velocity that results in a NR energy $E_{\rm{NR}}$. 
All input parameters to Eq.~\eqref{eq:drde} were set to the conventions given in Ref.~\cite{Baxter:2021pqo}.

Besides the pure NR process, we consider the NR-induced electron recoil (ER) signals by the Migdal effect~\cite{Ibe:2017yqa}, which has been employed by XENON, LZ, LUX, CRESST, SuperCDMS and CDEX experiments~\cite{XENON:2019zpr,LZ:2023poo,LUX:2018akb,CRESST:2019jnq,SuperCDMS:2023sql,CDEX:2021cll} to extend the reach for low-mass DM searches~\footnote{One should also be aware that there are ongoing efforts to directly measure the Migdal effect in liquid xenon using neutron calibration~\cite{Araujo:2022wjh,Xu:2023wev,bang2023migdal}. The findings are still contradictory, therefore the systematic uncertainty in the theoretical prediction remains to be settled.}. 
When a DM particle scatters with a xenon atom, the nucleus undergoes an abrupt momentum change with respect to the orbital electrons, resulting in the excitation or ionization of the atomic electrons due to the lack of transient movement of the electron cloud. This effect leads to the possible generation of ER signals in the keV range that accompany the primary NR. Therefore, even if the NR energy deposition is below the detection threshold, the ER energy deposition due to the Migdal effect can still be detected, providing a way to probe low-mass DM particles that are otherwise not detectable in PandaX. In this Letter, we only consider the ionization process, 
as the excitation probabilities are negligible in the energy region of interest~\cite{Cox:2022ekg, Ibe:2017yqa}.
The differential rate of ionization electron with energy $E_{\rm{ER}}$ is given by folding the NR spectrum in Eq.~\eqref{eq:drde} with the transition rate~\cite{Ibe:2017yqa},

\begin{equation}
\begin{split}
\label{eq:migdal_drde}
\frac{d R}{d E_{\mathrm{ER}}} & =\int d E_{\mathrm{NR}} d v \frac{d^2 R}{d E_{\mathrm{NR}} d v} \\
& \times \frac{1}{2 \pi} \sum_{n, l} \frac{d}{d E_{\mathrm{ER}}} p^{\mathrm{ion}}\left[n l \rightarrow\left(E_{\mathrm{ER}}-E_{n l}\right)\right],
\end{split}
\end{equation}

where $p^{\mathrm{ion}}\left[n l \rightarrow\left(E_{\mathrm{ER}}-E_{n l}\right)\right]$ is the probability for an atomic electron with quantum numbers $(n,l)$ and binding energy $E_{nl}$ to be ionized and receive a kinetic energy of $E_{\rm{ER}}-E_{nl}$. 
It is related to the electron momentum relative to the struck nucleus, therefore depends on $E_{\rm{NR}}$.
Similar to Refs.~\cite{XENON:2019zpr,LUX:2018akb}, we only take into account the contributions from the ionization of M-shell ($n$=3) and N-shell ($n$=4) electrons. The binding energies of inner shells ($n$=1, 2) are too strong to contribute significantly. 
The contribution from xenon valence electrons ($n$=5) has also been conservatively omitted, different from the treatment in Ref.~\cite{Qiao:2023pbw}, as the ambient atoms in the liquid may lead to large uncertainty in the ionization energy. 
We only consider DM masses up to 2~GeV, above which the contribution of NR signals becomes comparable or dominant.

In this Letter, the DM candidates are selected in two complementary ways, with and without requiring the presence of the $S1$ signal in each event. The former (denoted as $S1$-$S2$ below), has lower background contamination but a higher energy threshold, while the latter ($S2$-only) is the opposite.

\begin{figure}[t]
  \includegraphics[width=0.48\textwidth]{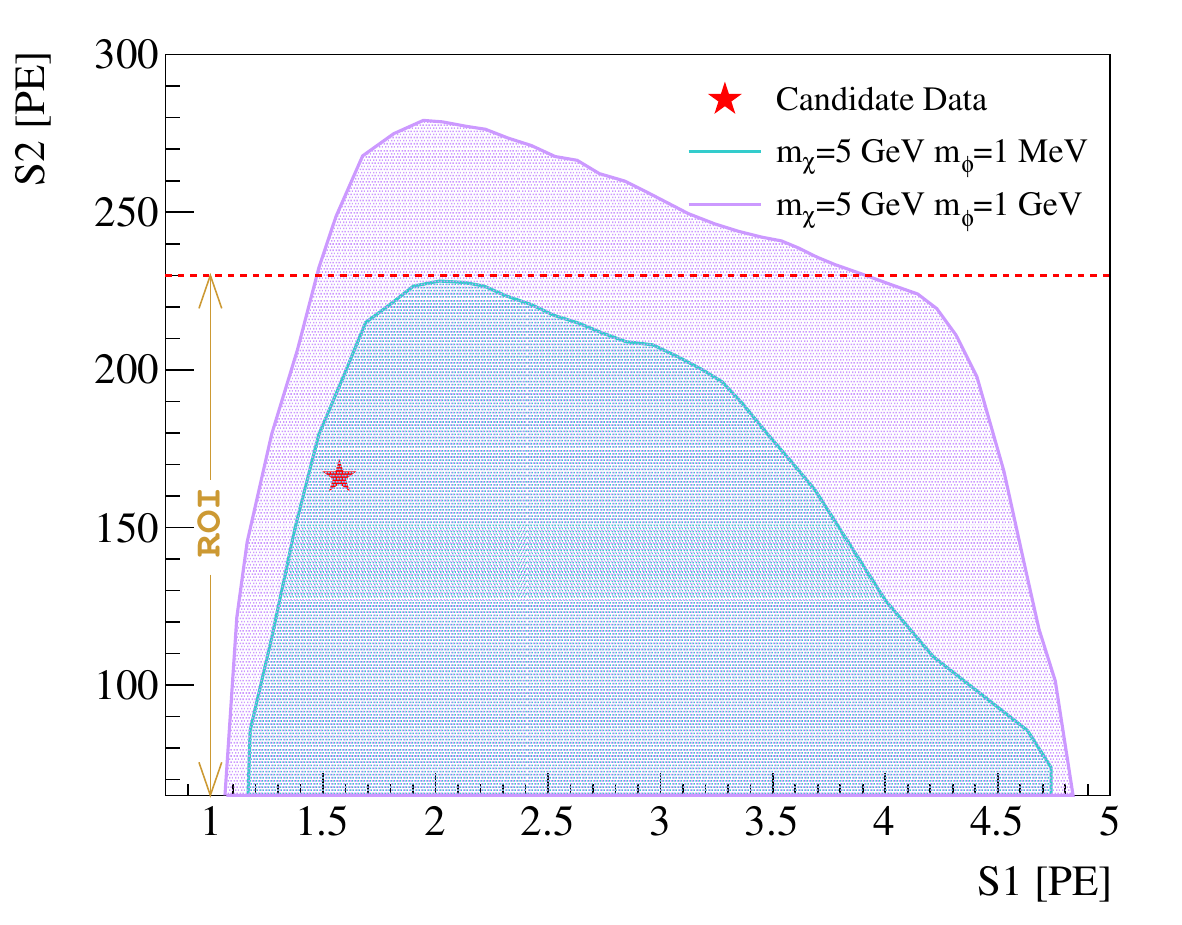}
  \caption{Selected DM candidate event in the $S1$-$S2$ plane after the BDT cut with the requirement of 2-hit $S1$.
  The cyan and violet lines represent the 95\% contour of the signals for a DM mass of 5~$\rm{GeV/c^2}$ with mediator masses of 1~$\rm{MeV/c^2}$ and 1~$\rm{GeV/c^2}$, respectively.
  }
  \label{fig:2hitdata}
\end{figure}

For the $S1$-$S2$ data, we follow the same procedure for the solar $^{8}$B neutrino and low-mass DM search as in Ref.~\cite{PandaX:2022aac}, using the commissioning data with an exposure of 0.48 tonne-year. Each event requires one pair of physically correlated $S1$ and $S2$ signals within the fiducial volume of the TPC. The $S1$ signal is required to have 2 or 3 coincident PMT hits (corresponding to a 1\%-acceptance threshold of 0.95~keV$_{\rm{nr}}$), and the $S2$ signal needs to be 65\textendash230 PE for 2-hit $S1$ and 65\textendash190 PE for 3-hit $S1$. A boosted decision tree (BDT) algorithm-based selection is applied to suppress the dominated accidental coincidence background from randomly paired $S1$ and $S2$ signals.  
After the BDT cut, there is one event left in the 2-hit $S1$ data (Fig.~\ref{fig:2hitdata}). The total number of background events, dominated by the accidental background and the $^{8}$B neutrino, is estimated to be \textcolor{black}{2.9\,$\pm$\,1.0}. No 3-hit $S1$ events survive and the background prediction is \textcolor{black}{ 0.4\,$\pm$\,0.1} events.

\begin{figure}[h]
  \includegraphics[width=0.48\textwidth]{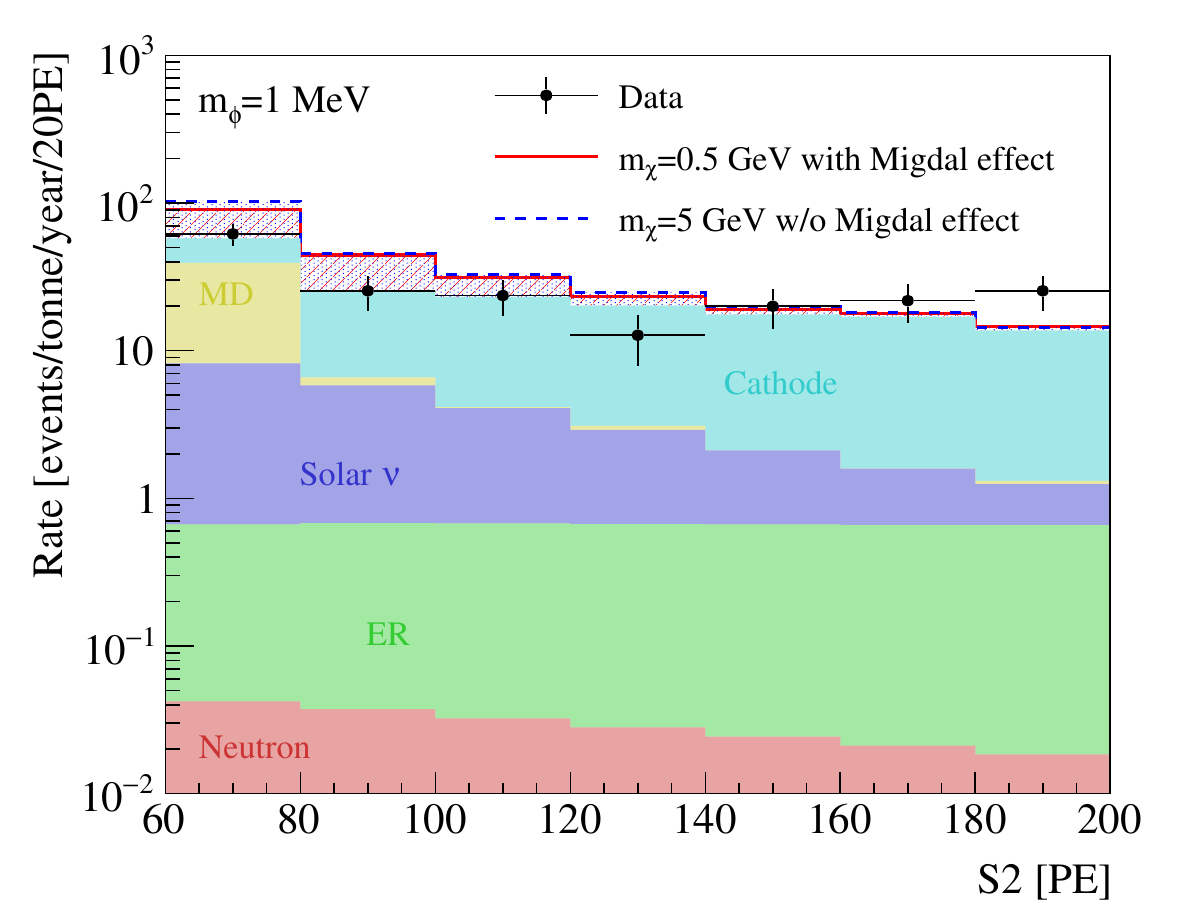}
  \caption{Distribution of selected candidates in S2-only data and background components. The expected signals in PandaX-4T with and without the Migdal effect for DM masses of 0.5 and 5~$\rm{GeV/c^2}$ are shown in red solid and blue dashed lines, respectively. In both cases, $m_{\phi}$ is set to be 1~$\rm{MeV/c^2}$, and zero-momentum-transfer DM-nucleon cross section of 3\,$\times\,10^{-38}$~cm$^2$ and 8\,$\times\,10^{-40}$~cm$^2$ (close to the later exclusion limits) are assumed, respectively.}
  \label{fig:US2data}
\end{figure}

For the $S2$-only data, we used the same datasets and procedure as in 
Ref.~\cite{PandaX:2022xqx} that correspond to an effective exposure of 0.55~tonne-year. 
The dataset, the background models, and the conversion response matrix from a given energy deposition to a distribution of $S2$ are provided in a public repository~\footnote{see \url{https://pandax.sjtu.edu.cn/public/data_release/PandaX-4T/run0_S2_only/}.}.
In total, there are 105 events with unpaired $S2$ (with no accompanying $S1$ greater than 2~PE) between 60 and 200~PE~\cite{PandaX:2022xqx}, and the detection energy threshold is approximately 0.8~keV$_{\rm{nr}}$ corresponding to a 14\% acceptance. 
The background is mainly composed of electrons from TPC cathode, microdischarge noises (MD) from the electrodes, and solar neutrinos. 
Figure~\ref{fig:US2data} shows the $S2$ distribution in data, in comparison with the background from the background-only best fit.

\begin{figure*}[ht]
    \includegraphics[width=0.48\textwidth]{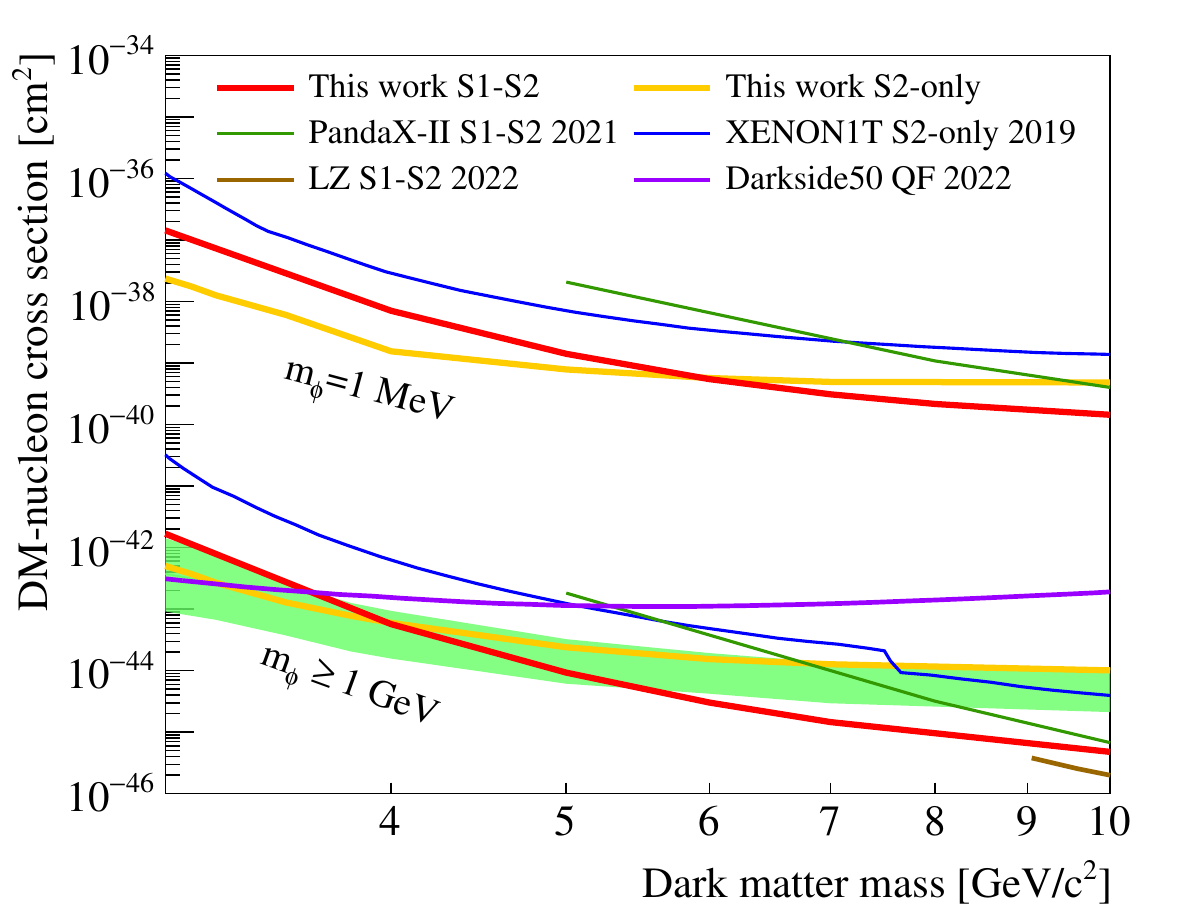}
    \includegraphics[width=0.48\textwidth]{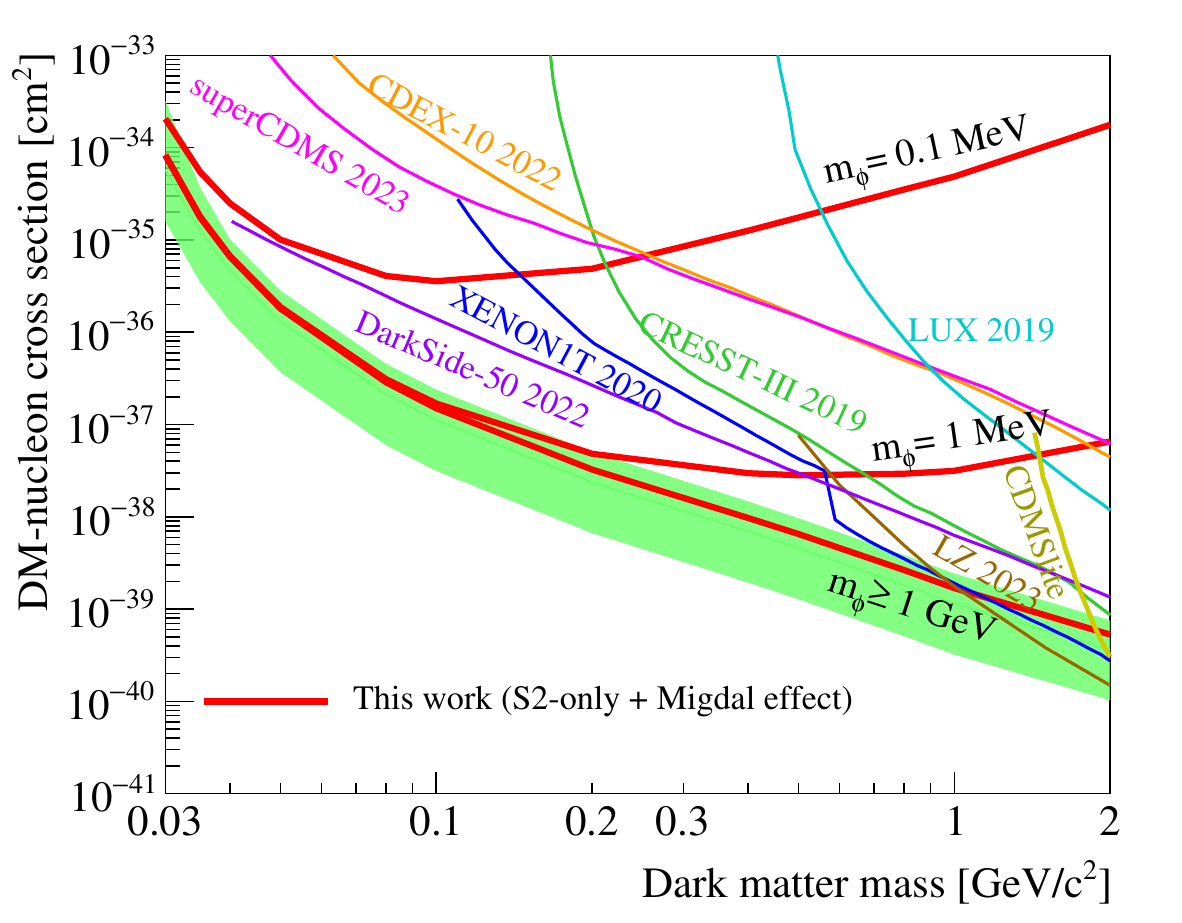}
    \caption{ {\it Left:} the 90\% C.L. upper limits on the zero-momentum DM-nucleon cross section for light mediator DM models for mediator masses 1~$\rm{MeV/c^2}$ and $\geq 1~\rm{GeV/c^2}$. The red lines represent the exclusion limit of this work using $S1$-$S2$ data. The orange lines and green shaded region represent the limits and $\pm$1$\sigma$ sensitivity (for $m_\phi$ $\geq 1~\rm{GeV/c^2}$) with $S2$-only data. For comparison, the limits from our previous analysis based on the full dataset of PandaX-II data release in 2021~\cite{PandaX-II:2021lap} (green) and the results of XENON1T $S2$-only~\cite{XENON:2019gfn} (blue), LZ $S1$-$S2$~\cite{LZ:2022lsv} (brown), and DarkSide-50~\cite{DarkSide-50:2022qzh} with quenching fluctuations (violet) are shown. 
    {\it Right:} upper limits on the DM-nucleon cross section at 90\% C.L. obtained with a signal including the Migdal effect for mediator masses 0.1 and 1~$\rm{MeV/c^2}$ and $\geq 1~\rm{GeV/c^2}$ (red lines), together with $\pm$1$\sigma$ sensitivity for $m_\phi$ $\geq 1~\rm{GeV/c^2}$ (green shaded area). Also shown are limits under the heavy mediator case from CRESST-III~\cite{CRESST:2019jnq} (green), XENON1T~\cite{XENON:2019zpr} (blue), LUX~\cite{LUX:2018akb} (cyan), LZ~\cite{LZ:2023poo} (brown), CDMSlite~\cite{SuperCDMS:2015eex} (yellow), SuperCDMS~\cite{SuperCDMS:2023sql} (magenta) , DarkSide-50~\cite{DarkSide:2022dhx} (violet), and CDEX-10~\cite{CDEX:2021cll} (orange).
    }
  \label{fig:limit_with3mphi}
\end{figure*}

For the statistical inference of DM signals, the binned profile likelihood ratio~\cite{Cowan:2010js,Cowan:2011an} is constructed as the test statistics, including the same treatments for the systematic uncertainties as in Refs.~\cite{PandaX:2022aac,PandaX:2022xqx}. The left panel of Fig.~\ref{fig:limit_with3mphi} shows the zero-momentum DM-nucleon cross section limits for mediator mass of 1~$\rm{MeV/c^2}$ or 1~$\rm{GeV/c^2}$, using both $S1$-$S2$ data and $S2$-only data for DM masses ranging from 3~$\rm{GeV/c^2}$ to 10~$\rm{GeV/c^2}$. As expected, the limits obtained from $S2$-only data are more stringent than those from $S1$-$S2$ data at smaller DM masses. At 5~$\rm{GeV/c^2}$, our limits are 1 order of magnitude stronger than the previous results from PandaX-II experiment. The right panel of Fig.~\ref{fig:limit_with3mphi} shows the limits using $S2$-only data with Migdal effect for DM masses between 30~$\rm{MeV/c^2}$ and 2~$\rm{GeV/c^2}$. 
The limits for $m_\phi$ = 1~$\rm{GeV/c^2}$ is within 1\% of the presented heavy mediator limit.
Our results represent a significant improvement over other experiments. For example, for DM masses ranging from 100 to 500~$\rm{MeV}$/$c^2$, our limits are 1 to 2 orders of magnitude stronger than those of XENON1T in the case of a heavy mediator. 
Note that we have assumed the so-called constant-W model for the charge yield all the way to zero energy, to be consistent with our earlier work in Refs.~\cite{PandaX-II:2021nsg,PandaX:2022xqx},  whereas in the XENON1T treatment in Ref.~\cite{XENON:2019zpr}, the charge yield was truncated below 186~$\rm{eV}$. This is partly responsible for our tighter constraint in this Letter. For the upper limits in this Letter, we have checked that the Earth attenuation effect~\cite{PandaX-II:2021kai} is negligible. For example, considering a scattering cross section of 10$^{-30}$~cm$^2$, along with $m_\phi$ at 0.1~$\rm{MeV/c^2}$ and $m_\chi$ at both 0.04 and 2~$\rm{GeV/c^2}$, the impact on event rates was found to be 1.6\% and 1.9\%, respectively.

\mbox{\textit{Constraints to dark-photon-mediated DM model.}\textemdash Next,}
we consider the dark mediator as a dark photon, 
with its kinetic mixing strength to ordinary photon denoted as $\epsilon_{\gamma}$.
Such a mixing provides an important connection between DM in the hidden sector and SM particles in the visible sector~\cite{essig2013dark,caputo2021dark,Hearty:2022wij,Fuyuto:2019vfe,cheng2022c} in the early Universe. 
One typical case is that the DM abundance is set by its annihilation into the SM particles, dominated by the process with an s-channel dark photon mediator~\cite{battaglieri2017us}. 
For scalar DM \footnote{For fermionic DM, the annihilation cross section is not suppressed by the velocity, and the model is ruled out by the Planck CMB data~\cite{Planck:2018vyg,battaglieri2017us}.}, the annihilation cross section scales as $\langle\sigma v\rangle \sim\frac{\epsilon_{\gamma}^2\alpha_{\chi}\alpha_{\rm EM} m^2_\chi v^2}{m^4_\phi}$ for a mediator with $m_\phi\gg2m_\chi$ and a negligible decay width~\cite{battaglieri2017us}. Here, $\alpha_{\rm EM}$ and $\alpha_\chi$ are the fine structure constants in the visible and dark sectors, respectively.

\begin{figure}[htb]
    \centering
    \includegraphics[width=0.48\textwidth]{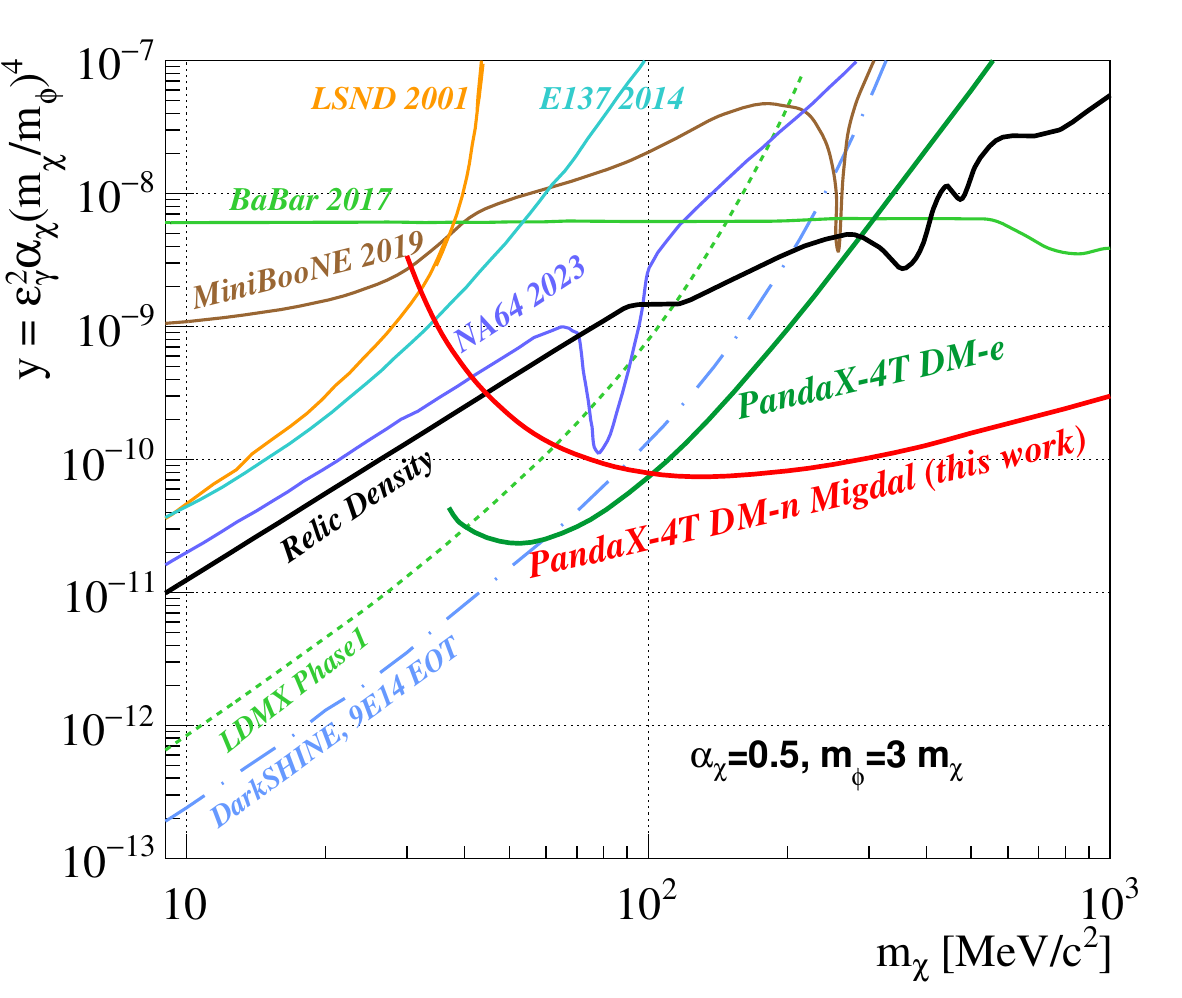}
    \caption{The 90\% C.L. exclusion limit (red) on the dimensionless interaction strength from PandaX-4T S2-only data for DM masses ranging from 30~$\rm{MeV/c^2}$ to 1~$\rm{GeV/c^2}$, assuming $m_\phi \gtrsim 0.05~m_\chi$. 
    The limit from PandaX-4T DM electron~\cite{PandaX:2022xqx} is also shown as the dark green curve. The black curve refers to the constraint from the observed relic DM density for the scalar elastic DM~\cite{battaglieri2017us}. Also included are results of NA64~\cite{NA64:2023wbi} (azure), MiniBooNE~\cite{MiniBooNEDM:2018cxm} (brown) , LSND~\cite{LSND:2001akn} (orange), E137~\cite{Batell:2014mga} (cyan), and $BABAR$~\cite{BaBar:2017tiz} (green) experiments, where $\alpha_{\chi}=0.5$ and $m_{\phi}=3~m_{\chi}$ are assumed.
    The projected limits from the LDMX experiment~\cite{Akesson:2022vza} (phase 1) and DarkSHINE experiment~\cite{Chen:2022liu} ($9\times10^{14}$ electron on target) are shown as the green dashed line and blue dot-dashed line, respectively. Note that exclusions and sensitivity curves from the dark photon experiments scale with $\alpha_{\chi}\times (m_{\chi}/m_{\phi})^4$.}
    \label{fig:migdal_ylimit}
\end{figure}

\noindent
As we can see, the aforementioned cross section depends on $m_\chi$ and a dimensionless parameter $y = \epsilon_{\gamma}^2\alpha_{\chi}(m_\chi/m_\phi)^4$ ~\cite{izaguirre1505accelerating}. The observed DM relic abundance requires $\langle\sigma v\rangle \sim \langle\sigma v\rangle_{\rm{relic}}$, which in turn fixes $y$ for a given $m_\chi$ (see Fig.~\ref{fig:migdal_ylimit}). 

On the other hand, the zero-momentum DM-nucleon cross section in direct detection experiments is given by~\cite{Kaplinghat:2013yxa}
\begin{equation}
  \sigma|_{q^2=0} = \frac{16\pi\alpha_{\rm
      EM}\alpha_{\chi}\mu^{2}_{p}}
         {m^{4}_{\phi}}\left[\frac{\epsilon_{\gamma}Z}{A}\right]^{2},
         \label{eq:sigma0}
\end{equation}
where $Z$ is the proton number of the xenon nuclei. Therefore, the cross section $\sigma|_{q^2=0}$ scales with $y\times\mu^{2}_{p}/m_\chi^{4}$, which provides a constraint in $y$ versus $m_\chi$.

It is worth noting that for direct detection experiment, when $m_\phi \gtrsim 0.05~m_\chi$, the $q^2$ contribution is negligible in Eq.~\eqref{eq:drde}.
For $m_{\chi}<1$~$\rm{GeV/c^2}$, our results using $S2$-only data with the Migdal effect and with $m_{\phi}=1$~$\rm{GeV/c^2}$ (in Fig.~\ref{fig:limit_with3mphi}, right) clearly satisfy this condition. 
Figure~\ref{fig:migdal_ylimit} illustrates the resulting upper limit on the parameter $y$. Therefore, our results provide a direct test of dark-photon-mediated DM as the thermal relic, and rule out the aforementioned scalar DM model in the mass range between 30~$\rm{MeV/c^2}$ and 1~$\rm{GeV/c^2}$. 
Similarly, recent constraints on DM-electron cross sections ($\si{F_{DM}}=1$) from PandaX-4T~\cite{PandaX:2022xqx} can be as well translated into limits on $y$ by the relation $y=\sigma_e\frac{m_\chi^4}{16\pi \alpha_{\rm{EM}} \mu_e^2}$~\cite{Emken:2019tni}, where $\mu_e$ is the DM-electron reduced mass, also overlaid on Fig.~\ref{fig:migdal_ylimit}. 
We display this limit separately, as it bears no theoretical uncertainty associated with the Migdal effect. The limit is even stronger than that from DM-nucleon interaction for DM masses ranging from 38 to 100~$\rm{MeV/c^2}$. 
On the other hand, dedicated fixed target experiments for dark photon search (e.g NA64~\cite{NA64:2023wbi}, LSND~\cite{LSND:2001akn}, E137~\cite{Batell:2014mga}) can set upper limits on $y$, but usually assuming a particular value of $\alpha_\chi$ and a fixed ratio between $m_\phi$ and $m_\chi$. 
Future experiments such as LDMX~\cite{Akesson:2022vza} and DarkSHINE~\cite{Chen:2022liu} are expected to probe low-mass region that has not yet been accessible by PandaX.

{\it Summary.}\textemdash
In summary, we present a search for the interactions between DM particles and nucleons via a dark mediator in the PandaX-4T experiment.
The analysis used both low-energy $S1$-$S2$ and $S2$-only data from the commissioning run.  The results from $S2$-only data with Migdal effect are further used to constrain the interaction strength between dark photon and ordinary photon. The result provides the most stringent constraint for the thermal scalar DM model, in which the s-channel dark photon decays into DM pairs in the early Universe, for the DM masses ranging from 30~$\rm{MeV/c^2}$ to 1~$\rm{GeV/c^2}$. 

\begin{acknowledgments}
    We would like to thank Kun Liu and Tran Van Que for their useful discussions. This project is supported in part by grants from the National Natural Science Foundation of China (No.~12090060, No.~12090061, No.~11875190, No.~12005131, No.~12105052, No.~11905128, No.~11925502, No.~11835005), a grant from Office of Science and Technology, Shanghai Municipal Government (Grant No.~22JC1410100), and Chinese postdoctoral Science Foundation (Grant No.~2021M700859). We thank Double First Class Plan of the Shanghai Jiao Tong University and the Tsung-Dao Lee Institute Experimental Platform Development Fund for support. We also thank the sponsorship from the Hongwen Foundation in Hong Kong, Tencent Foundation in China, and Yangyang Development Fund. Finally, we thank the China Jinping Underground Laboratory administration and the Yalong River Hydropower Development Company Ltd. for indispensable logistical support and other help. 
\end{acknowledgments}

\bibliographystyle{apsrev4-1}
\bibliography{reference.bib}

\begin{thebibliography}{58}%
\makeatletter
\providecommand \@ifxundefined [1]{%
 \@ifx{#1\undefined}
}%
\providecommand \@ifnum [1]{%
 \ifnum #1\expandafter \@firstoftwo
 \else \expandafter \@secondoftwo
 \fi
}%
\providecommand \@ifx [1]{%
 \ifx #1\expandafter \@firstoftwo
 \else \expandafter \@secondoftwo
 \fi
}%
\providecommand \natexlab [1]{#1}%
\providecommand \enquote  [1]{``#1''}%
\providecommand \bibnamefont  [1]{#1}%
\providecommand \bibfnamefont [1]{#1}%
\providecommand \citenamefont [1]{#1}%
\providecommand \href@noop [0]{\@secondoftwo}%
\providecommand \href [0]{\begingroup \@sanitize@url \@href}%
\providecommand \@href[1]{\@@startlink{#1}\@@href}%
\providecommand \@@href[1]{\endgroup#1\@@endlink}%
\providecommand \@sanitize@url [0]{\catcode `\\12\catcode `\$12\catcode
  `\&12\catcode `\#12\catcode `\^12\catcode `\_12\catcode `\%12\relax}%
\providecommand \@@startlink[1]{}%
\providecommand \@@endlink[0]{}%
\providecommand \url  [0]{\begingroup\@sanitize@url \@url }%
\providecommand \@url [1]{\endgroup\@href {#1}{\urlprefix }}%
\providecommand \urlprefix  [0]{URL }%
\providecommand \Eprint [0]{\href }%
\providecommand \doibase [0]{http://dx.doi.org/}%
\providecommand \selectlanguage [0]{\@gobble}%
\providecommand \bibinfo  [0]{\@secondoftwo}%
\providecommand \bibfield  [0]{\@secondoftwo}%
\providecommand \translation [1]{[#1]}%
\providecommand \BibitemOpen [0]{}%
\providecommand \bibitemStop [0]{}%
\providecommand \bibitemNoStop [0]{.\EOS\space}%
\providecommand \EOS [0]{\spacefactor3000\relax}%
\providecommand \BibitemShut  [1]{\csname bibitem#1\endcsname}%
\let\auto@bib@innerbib\@empty
\bibitem [{\citenamefont {Meng}\ \emph {et~al.}(2021)\citenamefont {Meng} \emph
  {et~al.}}]{PandaX-4T:2021bab}%
  \BibitemOpen
  \bibfield  {author} {\bibinfo {author} {\bibfnamefont {Y.}~\bibnamefont
  {Meng}} \emph {et~al.} (\bibinfo {collaboration} {PandaX-4T}),\ }\href
  {\doibase 10.1103/PhysRevLett.127.261802} {\bibfield  {journal} {\bibinfo
  {journal} {Phys. Rev. Lett.}\ }\textbf {\bibinfo {volume} {127}},\ \bibinfo
  {pages} {261802} (\bibinfo {year} {2021})},\ \Eprint
  {http://arxiv.org/abs/2107.13438} {arXiv:2107.13438 [hep-ex]} \BibitemShut
  {NoStop}%
\bibitem [{\citenamefont {Aalbers}\ \emph
  {et~al.}(2023{\natexlab{a}})\citenamefont {Aalbers} \emph
  {et~al.}}]{LZ:2022lsv}%
  \BibitemOpen
  \bibfield  {author} {\bibinfo {author} {\bibfnamefont {J.}~\bibnamefont
  {Aalbers}} \emph {et~al.} (\bibinfo {collaboration} {LZ}),\ }\href {\doibase
  10.1103/PhysRevLett.131.041002} {\bibfield  {journal} {\bibinfo  {journal}
  {Phys. Rev. Lett.}\ }\textbf {\bibinfo {volume} {131}},\ \bibinfo {pages}
  {041002} (\bibinfo {year} {2023}{\natexlab{a}})},\ \Eprint
  {http://arxiv.org/abs/2207.03764} {arXiv:2207.03764 [hep-ex]} \BibitemShut
  {NoStop}%
\bibitem [{\citenamefont {Aprile}\ \emph {et~al.}(2023)\citenamefont {Aprile}
  \emph {et~al.}}]{XENON:2023sxq}%
  \BibitemOpen
  \bibfield  {author} {\bibinfo {author} {\bibfnamefont {E.}~\bibnamefont
  {Aprile}} \emph {et~al.} (\bibinfo {collaboration} {XENON}),\ }\href@noop {}
  {\  (\bibinfo {year} {2023})},\ \Eprint {http://arxiv.org/abs/2303.14729}
  {arXiv:2303.14729 [hep-ex]} \BibitemShut {NoStop}%
\bibitem [{\citenamefont {Ren}\ \emph {et~al.}(2018)\citenamefont {Ren} \emph
  {et~al.}}]{Ren:2018gyx}%
  \BibitemOpen
  \bibfield  {author} {\bibinfo {author} {\bibfnamefont {X.}~\bibnamefont
  {Ren}} \emph {et~al.} (\bibinfo {collaboration} {PandaX-II}),\ }\href
  {\doibase 10.1103/PhysRevLett.121.021304} {\bibfield  {journal} {\bibinfo
  {journal} {Phys. Rev. Lett.}\ }\textbf {\bibinfo {volume} {121}},\ \bibinfo
  {pages} {021304} (\bibinfo {year} {2018})},\ \Eprint
  {http://arxiv.org/abs/1802.06912} {arXiv:1802.06912 [hep-ph]} \BibitemShut
  {NoStop}%
\bibitem [{\citenamefont {Yang}\ \emph {et~al.}(2021)\citenamefont {Yang} \emph
  {et~al.}}]{PandaX-II:2021lap}%
  \BibitemOpen
  \bibfield  {author} {\bibinfo {author} {\bibfnamefont {J.}~\bibnamefont
  {Yang}} \emph {et~al.} (\bibinfo {collaboration} {PandaX-II}),\ }\href
  {\doibase 10.1007/s11433-021-1740-2} {\bibfield  {journal} {\bibinfo
  {journal} {Sci. China Phys. Mech. Astron.}\ }\textbf {\bibinfo {volume}
  {64}},\ \bibinfo {pages} {111062} (\bibinfo {year} {2021})},\ \Eprint
  {http://arxiv.org/abs/2104.14724} {arXiv:2104.14724 [hep-ex]} \BibitemShut
  {NoStop}%
\bibitem [{\citenamefont {Ning}\ \emph
  {et~al.}(2023{\natexlab{a}})\citenamefont {Ning} \emph
  {et~al.}}]{PandaX:2023toi}%
  \BibitemOpen
  \bibfield  {author} {\bibinfo {author} {\bibfnamefont {X.}~\bibnamefont
  {Ning}} \emph {et~al.} (\bibinfo {collaboration} {PandaX}),\ }\href {\doibase
  10.1038/s41586-023-05982-0} {\bibfield  {journal} {\bibinfo  {journal}
  {Nature}\ }\textbf {\bibinfo {volume} {618}},\ \bibinfo {pages} {47}
  (\bibinfo {year} {2023}{\natexlab{a}})}\BibitemShut {NoStop}%
\bibitem [{\citenamefont {Ning}\ \emph
  {et~al.}(2023{\natexlab{b}})\citenamefont {Ning} \emph
  {et~al.}}]{PandaX:2023tfq}%
  \BibitemOpen
  \bibfield  {author} {\bibinfo {author} {\bibfnamefont {X.}~\bibnamefont
  {Ning}} \emph {et~al.} (\bibinfo {collaboration} {PandaX}),\ }\href@noop {}
  {\  (\bibinfo {year} {2023}{\natexlab{b}})},\ \Eprint
  {http://arxiv.org/abs/2301.03010} {arXiv:2301.03010 [hep-ex]} \BibitemShut
  {NoStop}%
\bibitem [{\citenamefont {Bennett}\ \emph {et~al.}(2006)\citenamefont {Bennett}
  \emph {et~al.}}]{Muong-2:2006rrc}%
  \BibitemOpen
  \bibfield  {author} {\bibinfo {author} {\bibfnamefont {G.~W.}\ \bibnamefont
  {Bennett}} \emph {et~al.} (\bibinfo {collaboration} {Muon g-2}),\ }\href
  {\doibase 10.1103/PhysRevD.73.072003} {\bibfield  {journal} {\bibinfo
  {journal} {Phys. Rev. D}\ }\textbf {\bibinfo {volume} {73}},\ \bibinfo
  {pages} {072003} (\bibinfo {year} {2006})},\ \Eprint
  {http://arxiv.org/abs/hep-ex/0602035} {arXiv:hep-ex/0602035} \BibitemShut
  {NoStop}%
\bibitem [{\citenamefont {Abi}\ \emph {et~al.}(2021)\citenamefont {Abi} \emph
  {et~al.}}]{Muong-2:2021ojo}%
  \BibitemOpen
  \bibfield  {author} {\bibinfo {author} {\bibfnamefont {B.}~\bibnamefont
  {Abi}} \emph {et~al.} (\bibinfo {collaboration} {Muon g-2}),\ }\href
  {\doibase 10.1103/PhysRevLett.126.141801} {\bibfield  {journal} {\bibinfo
  {journal} {Phys. Rev. Lett.}\ }\textbf {\bibinfo {volume} {126}},\ \bibinfo
  {pages} {141801} (\bibinfo {year} {2021})},\ \Eprint
  {http://arxiv.org/abs/2104.03281} {arXiv:2104.03281 [hep-ex]} \BibitemShut
  {NoStop}%
\bibitem [{\citenamefont {Krasznahorkay}\ \emph {et~al.}(2016)\citenamefont
  {Krasznahorkay} \emph {et~al.}}]{Krasznahorkay:2015iga}%
  \BibitemOpen
  \bibfield  {author} {\bibinfo {author} {\bibfnamefont {A.~J.}\ \bibnamefont
  {Krasznahorkay}} \emph {et~al.},\ }\href {\doibase
  10.1103/PhysRevLett.116.042501} {\bibfield  {journal} {\bibinfo  {journal}
  {Phys. Rev. Lett.}\ }\textbf {\bibinfo {volume} {116}},\ \bibinfo {pages}
  {042501} (\bibinfo {year} {2016})},\ \Eprint
  {http://arxiv.org/abs/1504.01527} {arXiv:1504.01527 [nucl-ex]} \BibitemShut
  {NoStop}%
\bibitem [{\citenamefont {Tulin}\ and\ \citenamefont
  {Yu}(2018)}]{Tulin:2017ara}%
  \BibitemOpen
  \bibfield  {author} {\bibinfo {author} {\bibfnamefont {S.}~\bibnamefont
  {Tulin}}\ and\ \bibinfo {author} {\bibfnamefont {H.-B.}\ \bibnamefont {Yu}},\
  }\href {\doibase 10.1016/j.physrep.2017.11.004} {\bibfield  {journal}
  {\bibinfo  {journal} {Phys. Rept.}\ }\textbf {\bibinfo {volume} {730}},\
  \bibinfo {pages} {1} (\bibinfo {year} {2018})},\ \Eprint
  {http://arxiv.org/abs/1705.02358} {arXiv:1705.02358 [hep-ph]} \BibitemShut
  {NoStop}%
\bibitem [{\citenamefont {Hearty}(2022)}]{Hearty:2022wij}%
  \BibitemOpen
  \bibfield  {author} {\bibinfo {author} {\bibfnamefont {C.}~\bibnamefont
  {Hearty}},\ }\href {\doibase 10.1088/1742-6596/2391/1/012011} {\bibfield
  {journal} {\bibinfo  {journal} {J. Phys. Conf. Ser.}\ }\textbf {\bibinfo
  {volume} {2391}},\ \bibinfo {pages} {012011} (\bibinfo {year}
  {2022})}\BibitemShut {NoStop}%
\bibitem [{\citenamefont {Battaglieri}\ \emph {et~al.}(2017)\citenamefont
  {Battaglieri}, \citenamefont {Belloni}, \citenamefont {Chou}, \citenamefont
  {Cushman}, \citenamefont {Echenard}, \citenamefont {Essig}, \citenamefont
  {Estrada}, \citenamefont {Feng}, \citenamefont {Flaugher}, \citenamefont
  {Fox} \emph {et~al.}}]{battaglieri2017us}%
  \BibitemOpen
  \bibfield  {author} {\bibinfo {author} {\bibfnamefont {M.}~\bibnamefont
  {Battaglieri}}, \bibinfo {author} {\bibfnamefont {A.}~\bibnamefont
  {Belloni}}, \bibinfo {author} {\bibfnamefont {A.}~\bibnamefont {Chou}},
  \bibinfo {author} {\bibfnamefont {P.}~\bibnamefont {Cushman}}, \bibinfo
  {author} {\bibfnamefont {B.}~\bibnamefont {Echenard}}, \bibinfo {author}
  {\bibfnamefont {R.}~\bibnamefont {Essig}}, \bibinfo {author} {\bibfnamefont
  {J.}~\bibnamefont {Estrada}}, \bibinfo {author} {\bibfnamefont {J.~L.}\
  \bibnamefont {Feng}}, \bibinfo {author} {\bibfnamefont {B.}~\bibnamefont
  {Flaugher}}, \bibinfo {author} {\bibfnamefont {P.~J.}\ \bibnamefont {Fox}},
  \emph {et~al.},\ }\href@noop {} {\bibfield  {journal} {\bibinfo  {journal}
  {arXiv preprint arXiv:1707.04591}\ } (\bibinfo {year} {2017})}\BibitemShut
  {NoStop}%
\bibitem [{\citenamefont {Andreev}\ \emph {et~al.}(2023)\citenamefont {Andreev}
  \emph {et~al.}}]{NA64:2023wbi}%
  \BibitemOpen
  \bibfield  {author} {\bibinfo {author} {\bibfnamefont {Y.~M.}\ \bibnamefont
  {Andreev}} \emph {et~al.} (\bibinfo {collaboration} {NA64}),\ }\href@noop {}
  {\  (\bibinfo {year} {2023})},\ \Eprint {http://arxiv.org/abs/2307.02404}
  {arXiv:2307.02404 [hep-ex]} \BibitemShut {NoStop}%
\bibitem [{\citenamefont {Aguilar-Arevalo}\ \emph {et~al.}(2018)\citenamefont
  {Aguilar-Arevalo} \emph {et~al.}}]{MiniBooNEDM:2018cxm}%
  \BibitemOpen
  \bibfield  {author} {\bibinfo {author} {\bibfnamefont {A.~A.}\ \bibnamefont
  {Aguilar-Arevalo}} \emph {et~al.} (\bibinfo {collaboration} {MiniBooNE DM}),\
  }\href {\doibase 10.1103/PhysRevD.98.112004} {\bibfield  {journal} {\bibinfo
  {journal} {Phys. Rev. D}\ }\textbf {\bibinfo {volume} {98}},\ \bibinfo
  {pages} {112004} (\bibinfo {year} {2018})},\ \Eprint
  {http://arxiv.org/abs/1807.06137} {arXiv:1807.06137 [hep-ex]} \BibitemShut
  {NoStop}%
\bibitem [{\citenamefont {Auerbach}\ \emph {et~al.}(2001)\citenamefont
  {Auerbach} \emph {et~al.}}]{LSND:2001akn}%
  \BibitemOpen
  \bibfield  {author} {\bibinfo {author} {\bibfnamefont {L.~B.}\ \bibnamefont
  {Auerbach}} \emph {et~al.} (\bibinfo {collaboration} {LSND}),\ }\href
  {\doibase 10.1103/PhysRevD.63.112001} {\bibfield  {journal} {\bibinfo
  {journal} {Phys. Rev. D}\ }\textbf {\bibinfo {volume} {63}},\ \bibinfo
  {pages} {112001} (\bibinfo {year} {2001})},\ \Eprint
  {http://arxiv.org/abs/hep-ex/0101039} {arXiv:hep-ex/0101039} \BibitemShut
  {NoStop}%
\bibitem [{\citenamefont {Batell}\ \emph {et~al.}(2014)\citenamefont {Batell},
  \citenamefont {Essig},\ and\ \citenamefont {Surujon}}]{Batell:2014mga}%
  \BibitemOpen
  \bibfield  {author} {\bibinfo {author} {\bibfnamefont {B.}~\bibnamefont
  {Batell}}, \bibinfo {author} {\bibfnamefont {R.}~\bibnamefont {Essig}}, \
  and\ \bibinfo {author} {\bibfnamefont {Z.}~\bibnamefont {Surujon}},\ }\href
  {\doibase 10.1103/PhysRevLett.113.171802} {\bibfield  {journal} {\bibinfo
  {journal} {Phys. Rev. Lett.}\ }\textbf {\bibinfo {volume} {113}},\ \bibinfo
  {pages} {171802} (\bibinfo {year} {2014})},\ \Eprint
  {http://arxiv.org/abs/1406.2698} {arXiv:1406.2698 [hep-ph]} \BibitemShut
  {NoStop}%
\bibitem [{\citenamefont {Lees}\ \emph {et~al.}(2017)\citenamefont {Lees} \emph
  {et~al.}}]{BaBar:2017tiz}%
  \BibitemOpen
  \bibfield  {author} {\bibinfo {author} {\bibfnamefont {J.~P.}\ \bibnamefont
  {Lees}} \emph {et~al.} (\bibinfo {collaboration} {BaBar}),\ }\href {\doibase
  10.1103/PhysRevLett.119.131804} {\bibfield  {journal} {\bibinfo  {journal}
  {Phys. Rev. Lett.}\ }\textbf {\bibinfo {volume} {119}},\ \bibinfo {pages}
  {131804} (\bibinfo {year} {2017})},\ \Eprint
  {http://arxiv.org/abs/1702.03327} {arXiv:1702.03327 [hep-ex]} \BibitemShut
  {NoStop}%
\bibitem [{\citenamefont {Berlin}\ \emph {et~al.}(2019)\citenamefont {Berlin},
  \citenamefont {Blinov}, \citenamefont {Krnjaic}, \citenamefont {Schuster},\
  and\ \citenamefont {Toro}}]{berlin2019dark}%
  \BibitemOpen
  \bibfield  {author} {\bibinfo {author} {\bibfnamefont {A.}~\bibnamefont
  {Berlin}}, \bibinfo {author} {\bibfnamefont {N.}~\bibnamefont {Blinov}},
  \bibinfo {author} {\bibfnamefont {G.}~\bibnamefont {Krnjaic}}, \bibinfo
  {author} {\bibfnamefont {P.}~\bibnamefont {Schuster}}, \ and\ \bibinfo
  {author} {\bibfnamefont {N.}~\bibnamefont {Toro}},\ }\href@noop {} {\bibfield
   {journal} {\bibinfo  {journal} {Physical Review D}\ }\textbf {\bibinfo
  {volume} {99}},\ \bibinfo {pages} {075001} (\bibinfo {year}
  {2019})}\BibitemShut {NoStop}%
\bibitem [{\citenamefont {Chen}\ \emph {et~al.}(2023)\citenamefont {Chen} \emph
  {et~al.}}]{Chen:2022liu}%
  \BibitemOpen
  \bibfield  {author} {\bibinfo {author} {\bibfnamefont {J.}~\bibnamefont
  {Chen}} \emph {et~al.},\ }\href {\doibase 10.1007/s11433-022-1983-8}
  {\bibfield  {journal} {\bibinfo  {journal} {Sci. China Phys. Mech. Astron.}\
  }\textbf {\bibinfo {volume} {66}},\ \bibinfo {pages} {211062} (\bibinfo
  {year} {2023})}\BibitemShut {NoStop}%
\bibitem [{\citenamefont {Yang}\ \emph {et~al.}(2022)\citenamefont {Yang},
  \citenamefont {Chen}, \citenamefont {He}, \citenamefont {Huang},
  \citenamefont {Huang}, \citenamefont {Liu}, \citenamefont {Ren},
  \citenamefont {Wang}, \citenamefont {Wang}, \citenamefont {Yan} \emph
  {et~al.}}]{yang2022readout}%
  \BibitemOpen
  \bibfield  {author} {\bibinfo {author} {\bibfnamefont {J.}~\bibnamefont
  {Yang}}, \bibinfo {author} {\bibfnamefont {X.}~\bibnamefont {Chen}}, \bibinfo
  {author} {\bibfnamefont {C.}~\bibnamefont {He}}, \bibinfo {author}
  {\bibfnamefont {D.}~\bibnamefont {Huang}}, \bibinfo {author} {\bibfnamefont
  {Y.}~\bibnamefont {Huang}}, \bibinfo {author} {\bibfnamefont
  {J.}~\bibnamefont {Liu}}, \bibinfo {author} {\bibfnamefont {X.}~\bibnamefont
  {Ren}}, \bibinfo {author} {\bibfnamefont {A.}~\bibnamefont {Wang}}, \bibinfo
  {author} {\bibfnamefont {M.}~\bibnamefont {Wang}}, \bibinfo {author}
  {\bibfnamefont {B.}~\bibnamefont {Yan}},  \emph {et~al.},\ }\href@noop {}
  {\bibfield  {journal} {\bibinfo  {journal} {Journal of Instrumentation}\
  }\textbf {\bibinfo {volume} {17}},\ \bibinfo {pages} {T02004} (\bibinfo
  {year} {2022})}\BibitemShut {NoStop}%
\bibitem [{Note1()}]{Note1}%
  \BibitemOpen
  \bibinfo {note} {See \protect \url
  {https://www.caen.it/products/v1725/}.}\BibitemShut {Stop}%
\bibitem [{\citenamefont {Savage}\ \emph {et~al.}(2009)\citenamefont {Savage},
  \citenamefont {Gelmini}, \citenamefont {Gondolo},\ and\ \citenamefont
  {Freese}}]{Savage:2008er}%
  \BibitemOpen
  \bibfield  {author} {\bibinfo {author} {\bibfnamefont {C.}~\bibnamefont
  {Savage}}, \bibinfo {author} {\bibfnamefont {G.}~\bibnamefont {Gelmini}},
  \bibinfo {author} {\bibfnamefont {P.}~\bibnamefont {Gondolo}}, \ and\
  \bibinfo {author} {\bibfnamefont {K.}~\bibnamefont {Freese}},\ }\href
  {\doibase 10.1088/1475-7516/2009/04/010} {\bibfield  {journal} {\bibinfo
  {journal} {JCAP}\ }\textbf {\bibinfo {volume} {04}},\ \bibinfo {pages} {010}
  (\bibinfo {year} {2009})},\ \Eprint {http://arxiv.org/abs/0808.3607}
  {arXiv:0808.3607 [astro-ph]} \BibitemShut {NoStop}%
\bibitem [{\citenamefont {Kaplinghat}\ \emph {et~al.}(2014)\citenamefont
  {Kaplinghat}, \citenamefont {Tulin},\ and\ \citenamefont
  {Yu}}]{Kaplinghat:2013yxa}%
  \BibitemOpen
  \bibfield  {author} {\bibinfo {author} {\bibfnamefont {M.}~\bibnamefont
  {Kaplinghat}}, \bibinfo {author} {\bibfnamefont {S.}~\bibnamefont {Tulin}}, \
  and\ \bibinfo {author} {\bibfnamefont {H.-B.}\ \bibnamefont {Yu}},\ }\href
  {\doibase 10.1103/PhysRevD.89.035009} {\bibfield  {journal} {\bibinfo
  {journal} {Phys. Rev.}\ }\textbf {\bibinfo {volume} {D89}},\ \bibinfo {pages}
  {035009} (\bibinfo {year} {2014})},\ \Eprint {http://arxiv.org/abs/1310.7945}
  {arXiv:1310.7945 [hep-ph]} \BibitemShut {NoStop}%
\bibitem [{\citenamefont {Baxter}\ \emph {et~al.}(2021)\citenamefont {Baxter}
  \emph {et~al.}}]{Baxter:2021pqo}%
  \BibitemOpen
  \bibfield  {author} {\bibinfo {author} {\bibfnamefont {D.}~\bibnamefont
  {Baxter}} \emph {et~al.},\ }\href@noop {} {\  (\bibinfo {year} {2021})},\
  \Eprint {http://arxiv.org/abs/2105.00599} {arXiv:2105.00599 [hep-ex]}
  \BibitemShut {NoStop}%
\bibitem [{\citenamefont {Ibe}\ \emph {et~al.}(2018)\citenamefont {Ibe},
  \citenamefont {Nakano}, \citenamefont {Shoji},\ and\ \citenamefont
  {Suzuki}}]{Ibe:2017yqa}%
  \BibitemOpen
  \bibfield  {author} {\bibinfo {author} {\bibfnamefont {M.}~\bibnamefont
  {Ibe}}, \bibinfo {author} {\bibfnamefont {W.}~\bibnamefont {Nakano}},
  \bibinfo {author} {\bibfnamefont {Y.}~\bibnamefont {Shoji}}, \ and\ \bibinfo
  {author} {\bibfnamefont {K.}~\bibnamefont {Suzuki}},\ }\href {\doibase
  10.1007/JHEP03(2018)194} {\bibfield  {journal} {\bibinfo  {journal} {JHEP}\
  }\textbf {\bibinfo {volume} {03}},\ \bibinfo {pages} {194} (\bibinfo {year}
  {2018})},\ \Eprint {http://arxiv.org/abs/1707.07258} {arXiv:1707.07258
  [hep-ph]} \BibitemShut {NoStop}%
\bibitem [{\citenamefont {Aprile}\ \emph
  {et~al.}(2019{\natexlab{a}})\citenamefont {Aprile} \emph
  {et~al.}}]{XENON:2019zpr}%
  \BibitemOpen
  \bibfield  {author} {\bibinfo {author} {\bibfnamefont {E.}~\bibnamefont
  {Aprile}} \emph {et~al.} (\bibinfo {collaboration} {XENON}),\ }\href
  {\doibase 10.1103/PhysRevLett.123.241803} {\bibfield  {journal} {\bibinfo
  {journal} {Phys. Rev. Lett.}\ }\textbf {\bibinfo {volume} {123}},\ \bibinfo
  {pages} {241803} (\bibinfo {year} {2019}{\natexlab{a}})},\ \Eprint
  {http://arxiv.org/abs/1907.12771} {arXiv:1907.12771 [hep-ex]} \BibitemShut
  {NoStop}%
\bibitem [{\citenamefont {Aalbers}\ \emph
  {et~al.}(2023{\natexlab{b}})\citenamefont {Aalbers} \emph
  {et~al.}}]{LZ:2023poo}%
  \BibitemOpen
  \bibfield  {author} {\bibinfo {author} {\bibfnamefont {J.}~\bibnamefont
  {Aalbers}} \emph {et~al.} (\bibinfo {collaboration} {LZ}),\ }\href@noop {} {\
   (\bibinfo {year} {2023}{\natexlab{b}})},\ \Eprint
  {http://arxiv.org/abs/2307.15753} {arXiv:2307.15753 [hep-ex]} \BibitemShut
  {NoStop}%
\bibitem [{\citenamefont {Akerib}\ \emph {et~al.}(2019)\citenamefont {Akerib}
  \emph {et~al.}}]{LUX:2018akb}%
  \BibitemOpen
  \bibfield  {author} {\bibinfo {author} {\bibfnamefont {D.~S.}\ \bibnamefont
  {Akerib}} \emph {et~al.} (\bibinfo {collaboration} {LUX}),\ }\href {\doibase
  10.1103/PhysRevLett.122.131301} {\bibfield  {journal} {\bibinfo  {journal}
  {Phys. Rev. Lett.}\ }\textbf {\bibinfo {volume} {122}},\ \bibinfo {pages}
  {131301} (\bibinfo {year} {2019})},\ \Eprint
  {http://arxiv.org/abs/1811.11241} {arXiv:1811.11241 [astro-ph.CO]}
  \BibitemShut {NoStop}%
\bibitem [{\citenamefont {Abdelhameed}\ \emph {et~al.}(2019)\citenamefont
  {Abdelhameed} \emph {et~al.}}]{CRESST:2019jnq}%
  \BibitemOpen
  \bibfield  {author} {\bibinfo {author} {\bibfnamefont {A.~H.}\ \bibnamefont
  {Abdelhameed}} \emph {et~al.} (\bibinfo {collaboration} {CRESST}),\ }\href
  {\doibase 10.1103/PhysRevD.100.102002} {\bibfield  {journal} {\bibinfo
  {journal} {Phys. Rev. D}\ }\textbf {\bibinfo {volume} {100}},\ \bibinfo
  {pages} {102002} (\bibinfo {year} {2019})},\ \Eprint
  {http://arxiv.org/abs/1904.00498} {arXiv:1904.00498 [astro-ph.CO]}
  \BibitemShut {NoStop}%
\bibitem [{\citenamefont {Albakry}\ \emph {et~al.}(2023)\citenamefont {Albakry}
  \emph {et~al.}}]{SuperCDMS:2023sql}%
  \BibitemOpen
  \bibfield  {author} {\bibinfo {author} {\bibfnamefont {M.~F.}\ \bibnamefont
  {Albakry}} \emph {et~al.} (\bibinfo {collaboration} {SuperCDMS}),\
  }\href@noop {} {\  (\bibinfo {year} {2023})},\ \Eprint
  {http://arxiv.org/abs/2302.09115} {arXiv:2302.09115 [hep-ex]} \BibitemShut
  {NoStop}%
\bibitem [{\citenamefont {Liu}\ \emph {et~al.}(2022)\citenamefont {Liu} \emph
  {et~al.}}]{CDEX:2021cll}%
  \BibitemOpen
  \bibfield  {author} {\bibinfo {author} {\bibfnamefont {Z.~Z.}\ \bibnamefont
  {Liu}} \emph {et~al.} (\bibinfo {collaboration} {CDEX}),\ }\href {\doibase
  10.1103/PhysRevD.105.052005} {\bibfield  {journal} {\bibinfo  {journal}
  {Phys. Rev. D}\ }\textbf {\bibinfo {volume} {105}},\ \bibinfo {pages}
  {052005} (\bibinfo {year} {2022})},\ \Eprint
  {http://arxiv.org/abs/2111.11243} {arXiv:2111.11243 [hep-ex]} \BibitemShut
  {NoStop}%
\bibitem [{Note2()}]{Note2}%
  \BibitemOpen
  \bibinfo {note} {One should also be aware that there are ongoing efforts to
  directly measure the Migdal effect in liquid xenon using neutron
  calibration~\cite {Araujo:2022wjh,Xu:2023wev,bang2023migdal}. The findings
  are still contradictory, therefore the systematic uncertainty in the
  theoretical prediction remains to be settled.}\BibitemShut {Stop}%
\bibitem [{\citenamefont {Cox}\ \emph {et~al.}(2023)\citenamefont {Cox},
  \citenamefont {Dolan}, \citenamefont {McCabe},\ and\ \citenamefont
  {Quiney}}]{Cox:2022ekg}%
  \BibitemOpen
  \bibfield  {author} {\bibinfo {author} {\bibfnamefont {P.}~\bibnamefont
  {Cox}}, \bibinfo {author} {\bibfnamefont {M.~J.}\ \bibnamefont {Dolan}},
  \bibinfo {author} {\bibfnamefont {C.}~\bibnamefont {McCabe}}, \ and\ \bibinfo
  {author} {\bibfnamefont {H.~M.}\ \bibnamefont {Quiney}},\ }\href {\doibase
  10.1103/PhysRevD.107.035032} {\bibfield  {journal} {\bibinfo  {journal}
  {Phys. Rev. D}\ }\textbf {\bibinfo {volume} {107}},\ \bibinfo {pages}
  {035032} (\bibinfo {year} {2023})},\ \Eprint
  {http://arxiv.org/abs/2208.12222} {arXiv:2208.12222 [hep-ph]} \BibitemShut
  {NoStop}%
\bibitem [{\citenamefont {Qiao}\ \emph {et~al.}(2023)\citenamefont {Qiao},
  \citenamefont {Xia},\ and\ \citenamefont {Zhou}}]{Qiao:2023pbw}%
  \BibitemOpen
  \bibfield  {author} {\bibinfo {author} {\bibfnamefont {M.}~\bibnamefont
  {Qiao}}, \bibinfo {author} {\bibfnamefont {C.}~\bibnamefont {Xia}}, \ and\
  \bibinfo {author} {\bibfnamefont {Y.-F.}\ \bibnamefont {Zhou}},\ }\href@noop
  {} {\  (\bibinfo {year} {2023})},\ \Eprint {http://arxiv.org/abs/2307.12820}
  {arXiv:2307.12820 [hep-ph]} \BibitemShut {NoStop}%
\bibitem [{\citenamefont {Ma}\ \emph {et~al.}(2023)\citenamefont {Ma} \emph
  {et~al.}}]{PandaX:2022aac}%
  \BibitemOpen
  \bibfield  {author} {\bibinfo {author} {\bibfnamefont {W.}~\bibnamefont {Ma}}
  \emph {et~al.} (\bibinfo {collaboration} {PandaX}),\ }\href {\doibase
  10.1103/PhysRevLett.130.021802} {\bibfield  {journal} {\bibinfo  {journal}
  {Phys. Rev. Lett.}\ }\textbf {\bibinfo {volume} {130}},\ \bibinfo {pages}
  {021802} (\bibinfo {year} {2023})},\ \Eprint
  {http://arxiv.org/abs/2207.04883} {arXiv:2207.04883 [hep-ex]} \BibitemShut
  {NoStop}%
\bibitem [{\citenamefont {Li}\ \emph {et~al.}(2023)\citenamefont {Li} \emph
  {et~al.}}]{PandaX:2022xqx}%
  \BibitemOpen
  \bibfield  {author} {\bibinfo {author} {\bibfnamefont {S.}~\bibnamefont {Li}}
  \emph {et~al.} (\bibinfo {collaboration} {PandaX}),\ }\href {\doibase
  10.1103/PhysRevLett.130.261001} {\bibfield  {journal} {\bibinfo  {journal}
  {Phys. Rev. Lett.}\ }\textbf {\bibinfo {volume} {130}},\ \bibinfo {pages}
  {261001} (\bibinfo {year} {2023})},\ \Eprint
  {http://arxiv.org/abs/2212.10067} {arXiv:2212.10067 [hep-ex]} \BibitemShut
  {NoStop}%
\bibitem [{Note3()}]{Note3}%
  \BibitemOpen
  \bibinfo {note} {See \protect \url
  {https://pandax.sjtu.edu.cn/public/data_release/PandaX-4T/run0_S2_only/}.}\BibitemShut
  {Stop}%
\bibitem [{\citenamefont {Aprile}\ \emph
  {et~al.}(2019{\natexlab{b}})\citenamefont {Aprile} \emph
  {et~al.}}]{XENON:2019gfn}%
  \BibitemOpen
  \bibfield  {author} {\bibinfo {author} {\bibfnamefont {E.}~\bibnamefont
  {Aprile}} \emph {et~al.} (\bibinfo {collaboration} {XENON}),\ }\href
  {\doibase 10.1103/PhysRevLett.123.251801} {\bibfield  {journal} {\bibinfo
  {journal} {Phys. Rev. Lett.}\ }\textbf {\bibinfo {volume} {123}},\ \bibinfo
  {pages} {251801} (\bibinfo {year} {2019}{\natexlab{b}})},\ \Eprint
  {http://arxiv.org/abs/1907.11485} {arXiv:1907.11485 [hep-ex]} \BibitemShut
  {NoStop}%
\bibitem [{\citenamefont {Agnes}\ \emph
  {et~al.}(2023{\natexlab{a}})\citenamefont {Agnes} \emph
  {et~al.}}]{DarkSide-50:2022qzh}%
  \BibitemOpen
  \bibfield  {author} {\bibinfo {author} {\bibfnamefont {P.}~\bibnamefont
  {Agnes}} \emph {et~al.} (\bibinfo {collaboration} {DarkSide-50}),\ }\href
  {\doibase 10.1103/PhysRevD.107.063001} {\bibfield  {journal} {\bibinfo
  {journal} {Phys. Rev. D}\ }\textbf {\bibinfo {volume} {107}},\ \bibinfo
  {pages} {063001} (\bibinfo {year} {2023}{\natexlab{a}})},\ \Eprint
  {http://arxiv.org/abs/2207.11966} {arXiv:2207.11966 [hep-ex]} \BibitemShut
  {NoStop}%
\bibitem [{\citenamefont {Agnese}\ \emph {et~al.}(2016)\citenamefont {Agnese}
  \emph {et~al.}}]{SuperCDMS:2015eex}%
  \BibitemOpen
  \bibfield  {author} {\bibinfo {author} {\bibfnamefont {R.}~\bibnamefont
  {Agnese}} \emph {et~al.} (\bibinfo {collaboration} {SuperCDMS}),\ }\href
  {\doibase 10.1103/PhysRevLett.116.071301} {\bibfield  {journal} {\bibinfo
  {journal} {Phys. Rev. Lett.}\ }\textbf {\bibinfo {volume} {116}},\ \bibinfo
  {pages} {071301} (\bibinfo {year} {2016})},\ \Eprint
  {http://arxiv.org/abs/1509.02448} {arXiv:1509.02448 [astro-ph.CO]}
  \BibitemShut {NoStop}%
\bibitem [{\citenamefont {Agnes}\ \emph
  {et~al.}(2023{\natexlab{b}})\citenamefont {Agnes} \emph
  {et~al.}}]{DarkSide:2022dhx}%
  \BibitemOpen
  \bibfield  {author} {\bibinfo {author} {\bibfnamefont {P.}~\bibnamefont
  {Agnes}} \emph {et~al.} (\bibinfo {collaboration} {DarkSide}),\ }\href
  {\doibase 10.1103/PhysRevLett.130.101001} {\bibfield  {journal} {\bibinfo
  {journal} {Phys. Rev. Lett.}\ }\textbf {\bibinfo {volume} {130}},\ \bibinfo
  {pages} {101001} (\bibinfo {year} {2023}{\natexlab{b}})},\ \Eprint
  {http://arxiv.org/abs/2207.11967} {arXiv:2207.11967 [hep-ex]} \BibitemShut
  {NoStop}%
\bibitem [{\citenamefont {Cowan}\ \emph
  {et~al.}(2011{\natexlab{a}})\citenamefont {Cowan}, \citenamefont {Cranmer},
  \citenamefont {Gross},\ and\ \citenamefont {Vitells}}]{Cowan:2010js}%
  \BibitemOpen
  \bibfield  {author} {\bibinfo {author} {\bibfnamefont {G.}~\bibnamefont
  {Cowan}}, \bibinfo {author} {\bibfnamefont {K.}~\bibnamefont {Cranmer}},
  \bibinfo {author} {\bibfnamefont {E.}~\bibnamefont {Gross}}, \ and\ \bibinfo
  {author} {\bibfnamefont {O.}~\bibnamefont {Vitells}},\ }\href {\doibase
  10.1140/epjc/s10052-011-1554-0} {\bibfield  {journal} {\bibinfo  {journal}
  {Eur. Phys. J. C}\ }\textbf {\bibinfo {volume} {71}},\ \bibinfo {pages}
  {1554} (\bibinfo {year} {2011}{\natexlab{a}})},\ \bibinfo {note} {[Erratum:
  Eur.Phys.J.C 73, 2501 (2013)]},\ \Eprint {http://arxiv.org/abs/1007.1727}
  {arXiv:1007.1727 [physics.data-an]} \BibitemShut {NoStop}%
\bibitem [{\citenamefont {Cowan}\ \emph
  {et~al.}(2011{\natexlab{b}})\citenamefont {Cowan}, \citenamefont {Cranmer},
  \citenamefont {Gross},\ and\ \citenamefont {Vitells}}]{Cowan:2011an}%
  \BibitemOpen
  \bibfield  {author} {\bibinfo {author} {\bibfnamefont {G.}~\bibnamefont
  {Cowan}}, \bibinfo {author} {\bibfnamefont {K.}~\bibnamefont {Cranmer}},
  \bibinfo {author} {\bibfnamefont {E.}~\bibnamefont {Gross}}, \ and\ \bibinfo
  {author} {\bibfnamefont {O.}~\bibnamefont {Vitells}},\ }\href@noop {} {\
  (\bibinfo {year} {2011}{\natexlab{b}})},\ \Eprint
  {http://arxiv.org/abs/1105.3166} {arXiv:1105.3166 [physics.data-an]}
  \BibitemShut {NoStop}%
\bibitem [{\citenamefont {Cheng}\ \emph {et~al.}(2021)\citenamefont {Cheng}
  \emph {et~al.}}]{PandaX-II:2021nsg}%
  \BibitemOpen
  \bibfield  {author} {\bibinfo {author} {\bibfnamefont {C.}~\bibnamefont
  {Cheng}} \emph {et~al.} (\bibinfo {collaboration} {PandaX-II}),\ }\href
  {\doibase 10.1103/PhysRevLett.126.211803} {\bibfield  {journal} {\bibinfo
  {journal} {Phys. Rev. Lett.}\ }\textbf {\bibinfo {volume} {126}},\ \bibinfo
  {pages} {211803} (\bibinfo {year} {2021})},\ \Eprint
  {http://arxiv.org/abs/2101.07479} {arXiv:2101.07479 [hep-ex]} \BibitemShut
  {NoStop}%
\bibitem [{\citenamefont {Cui}\ \emph {et~al.}(2022)\citenamefont {Cui} \emph
  {et~al.}}]{PandaX-II:2021kai}%
  \BibitemOpen
  \bibfield  {author} {\bibinfo {author} {\bibfnamefont {X.}~\bibnamefont
  {Cui}} \emph {et~al.} (\bibinfo {collaboration} {PandaX-II}),\ }\href
  {\doibase 10.1103/PhysRevLett.128.171801} {\bibfield  {journal} {\bibinfo
  {journal} {Phys. Rev. Lett.}\ }\textbf {\bibinfo {volume} {128}},\ \bibinfo
  {pages} {171801} (\bibinfo {year} {2022})},\ \Eprint
  {http://arxiv.org/abs/2112.08957} {arXiv:2112.08957 [hep-ex]} \BibitemShut
  {NoStop}%
\bibitem [{\citenamefont {Essig}\ \emph {et~al.}(2013)\citenamefont {Essig},
  \citenamefont {Jaros}, \citenamefont {Wester}, \citenamefont {Adrian},
  \citenamefont {Andreas}, \citenamefont {Averett}, \citenamefont {Baker},
  \citenamefont {Batell}, \citenamefont {Battaglieri}, \citenamefont {Beacham}
  \emph {et~al.}}]{essig2013dark}%
  \BibitemOpen
  \bibfield  {author} {\bibinfo {author} {\bibfnamefont {R.}~\bibnamefont
  {Essig}}, \bibinfo {author} {\bibfnamefont {J.~A.}\ \bibnamefont {Jaros}},
  \bibinfo {author} {\bibfnamefont {W.}~\bibnamefont {Wester}}, \bibinfo
  {author} {\bibfnamefont {P.~H.}\ \bibnamefont {Adrian}}, \bibinfo {author}
  {\bibfnamefont {S.}~\bibnamefont {Andreas}}, \bibinfo {author} {\bibfnamefont
  {T.}~\bibnamefont {Averett}}, \bibinfo {author} {\bibfnamefont
  {O.}~\bibnamefont {Baker}}, \bibinfo {author} {\bibfnamefont
  {B.}~\bibnamefont {Batell}}, \bibinfo {author} {\bibfnamefont
  {M.}~\bibnamefont {Battaglieri}}, \bibinfo {author} {\bibfnamefont
  {J.}~\bibnamefont {Beacham}},  \emph {et~al.},\ }\href@noop {} {\bibfield
  {journal} {\bibinfo  {journal} {arXiv preprint arXiv:1311.0029}\ } (\bibinfo
  {year} {2013})}\BibitemShut {NoStop}%
\bibitem [{\citenamefont {Caputo}\ \emph {et~al.}(2021)\citenamefont {Caputo},
  \citenamefont {Millar}, \citenamefont {O’Hare},\ and\ \citenamefont
  {Vitagliano}}]{caputo2021dark}%
  \BibitemOpen
  \bibfield  {author} {\bibinfo {author} {\bibfnamefont {A.}~\bibnamefont
  {Caputo}}, \bibinfo {author} {\bibfnamefont {A.~J.}\ \bibnamefont {Millar}},
  \bibinfo {author} {\bibfnamefont {C.~A.}\ \bibnamefont {O’Hare}}, \ and\
  \bibinfo {author} {\bibfnamefont {E.}~\bibnamefont {Vitagliano}},\
  }\href@noop {} {\bibfield  {journal} {\bibinfo  {journal} {Physical Review
  D}\ }\textbf {\bibinfo {volume} {104}},\ \bibinfo {pages} {095029} (\bibinfo
  {year} {2021})}\BibitemShut {NoStop}%
\bibitem [{\citenamefont {Fuyuto}\ \emph {et~al.}(2020)\citenamefont {Fuyuto},
  \citenamefont {He}, \citenamefont {Li},\ and\ \citenamefont
  {Ramsey-Musolf}}]{Fuyuto:2019vfe}%
  \BibitemOpen
  \bibfield  {author} {\bibinfo {author} {\bibfnamefont {K.}~\bibnamefont
  {Fuyuto}}, \bibinfo {author} {\bibfnamefont {X.-G.}\ \bibnamefont {He}},
  \bibinfo {author} {\bibfnamefont {G.}~\bibnamefont {Li}}, \ and\ \bibinfo
  {author} {\bibfnamefont {M.}~\bibnamefont {Ramsey-Musolf}},\ }\href {\doibase
  10.1103/PhysRevD.101.075016} {\bibfield  {journal} {\bibinfo  {journal}
  {Phys. Rev. D}\ }\textbf {\bibinfo {volume} {101}},\ \bibinfo {pages}
  {075016} (\bibinfo {year} {2020})},\ \Eprint
  {http://arxiv.org/abs/1902.10340} {arXiv:1902.10340 [hep-ph]} \BibitemShut
  {NoStop}%
\bibitem [{\citenamefont {Cheng}\ \emph {et~al.}(2022)\citenamefont {Cheng},
  \citenamefont {He}, \citenamefont {Ramsey-Musolf},\ and\ \citenamefont
  {Sun}}]{cheng2022c}%
  \BibitemOpen
  \bibfield  {author} {\bibinfo {author} {\bibfnamefont {Y.}~\bibnamefont
  {Cheng}}, \bibinfo {author} {\bibfnamefont {X.-G.}\ \bibnamefont {He}},
  \bibinfo {author} {\bibfnamefont {M.~J.}\ \bibnamefont {Ramsey-Musolf}}, \
  and\ \bibinfo {author} {\bibfnamefont {J.}~\bibnamefont {Sun}},\ }\href@noop
  {} {\bibfield  {journal} {\bibinfo  {journal} {Physical Review D}\ }\textbf
  {\bibinfo {volume} {105}},\ \bibinfo {pages} {095010} (\bibinfo {year}
  {2022})}\BibitemShut {NoStop}%
\bibitem [{Note4()}]{Note4}%
  \BibitemOpen
  \bibinfo {note} {For fermionic DM, the annihilation cross section is not
  suppressed by the velocity, and the model is ruled out by the Planck CMB
  data~\cite {Planck:2018vyg,battaglieri2017us}.}\BibitemShut {Stop}%
\bibitem [{\citenamefont {\r{A}kesson}\ \emph {et~al.}(2022)\citenamefont
  {\r{A}kesson} \emph {et~al.}}]{Akesson:2022vza}%
  \BibitemOpen
  \bibfield  {author} {\bibinfo {author} {\bibfnamefont {T.}~\bibnamefont
  {\r{A}kesson}} \emph {et~al.},\ }in\ \href@noop {} {\emph {\bibinfo
  {booktitle} {{Snowmass 2021}}}}\ (\bibinfo {year} {2022})\ \Eprint
  {http://arxiv.org/abs/2203.08192} {arXiv:2203.08192 [hep-ex]} \BibitemShut
  {NoStop}%
\bibitem [{\citenamefont {Izaguirre}\ \emph {et~al.}()\citenamefont
  {Izaguirre}, \citenamefont {Krnjaic}, \citenamefont {Schuster},\ and\
  \citenamefont {Toro}}]{izaguirre1505accelerating}%
  \BibitemOpen
  \bibfield  {author} {\bibinfo {author} {\bibfnamefont {E.}~\bibnamefont
  {Izaguirre}}, \bibinfo {author} {\bibfnamefont {G.}~\bibnamefont {Krnjaic}},
  \bibinfo {author} {\bibfnamefont {P.}~\bibnamefont {Schuster}}, \ and\
  \bibinfo {author} {\bibfnamefont {N.}~\bibnamefont {Toro}},\ }\href@noop {}
  {\bibinfo  {journal} {arXiv preprint arXiv:1505.00011}\ }\BibitemShut
  {NoStop}%
\bibitem [{\citenamefont {Emken}\ \emph {et~al.}(2019)\citenamefont {Emken},
  \citenamefont {Essig}, \citenamefont {Kouvaris},\ and\ \citenamefont
  {Sholapurkar}}]{Emken:2019tni}%
  \BibitemOpen
\bibfield  {journal} {  }\bibfield  {author} {\bibinfo {author} {\bibfnamefont
  {T.}~\bibnamefont {Emken}}, \bibinfo {author} {\bibfnamefont
  {R.}~\bibnamefont {Essig}}, \bibinfo {author} {\bibfnamefont
  {C.}~\bibnamefont {Kouvaris}}, \ and\ \bibinfo {author} {\bibfnamefont
  {M.}~\bibnamefont {Sholapurkar}},\ }\href {\doibase
  10.1088/1475-7516/2019/09/070} {\bibfield  {journal} {\bibinfo  {journal}
  {JCAP}\ }\textbf {\bibinfo {volume} {09}},\ \bibinfo {pages} {070} (\bibinfo
  {year} {2019})},\ \Eprint {http://arxiv.org/abs/1905.06348} {arXiv:1905.06348
  [hep-ph]} \BibitemShut {NoStop}%
\bibitem [{\citenamefont {Ara\'ujo}\ \emph {et~al.}(2023)\citenamefont
  {Ara\'ujo} \emph {et~al.}}]{Araujo:2022wjh}%
  \BibitemOpen
  \bibfield  {author} {\bibinfo {author} {\bibfnamefont {H.~M.}\ \bibnamefont
  {Ara\'ujo}} \emph {et~al.},\ }\href {\doibase
  10.1016/j.astropartphys.2023.102853} {\bibfield  {journal} {\bibinfo
  {journal} {Astropart. Phys.}\ }\textbf {\bibinfo {volume} {151}},\ \bibinfo
  {pages} {102853} (\bibinfo {year} {2023})},\ \Eprint
  {http://arxiv.org/abs/2207.08284} {arXiv:2207.08284 [hep-ex]} \BibitemShut
  {NoStop}%
\bibitem [{\citenamefont {Xu}\ \emph {et~al.}(2023)\citenamefont {Xu} \emph
  {et~al.}}]{Xu:2023wev}%
  \BibitemOpen
  \bibfield  {author} {\bibinfo {author} {\bibfnamefont {J.}~\bibnamefont {Xu}}
  \emph {et~al.},\ }\href@noop {} {\  (\bibinfo {year} {2023})},\ \Eprint
  {http://arxiv.org/abs/2307.12952} {arXiv:2307.12952 [hep-ex]} \BibitemShut
  {NoStop}%
\bibitem [{\citenamefont {Bang}\ \emph {et~al.}(2023)\citenamefont {Bang},
  \citenamefont {Vaitkus},\ and\ \citenamefont {Ding}}]{bang2023migdal}%
  \BibitemOpen
  \bibfield  {author} {\bibinfo {author} {\bibfnamefont {J.}~\bibnamefont
  {Bang}}, \bibinfo {author} {\bibfnamefont {A.}~\bibnamefont {Vaitkus}}, \
  and\ \bibinfo {author} {\bibfnamefont {C.}~\bibnamefont {Ding}},\ }\href@noop
  {} {\bibfield  {journal} {\bibinfo  {journal} {Bulletin of the American
  Physical Society}\ } (\bibinfo {year} {2023})}\BibitemShut {NoStop}%
\bibitem [{\citenamefont {Aghanim}\ \emph {et~al.}(2020)\citenamefont {Aghanim}
  \emph {et~al.}}]{Planck:2018vyg}%
  \BibitemOpen
  \bibfield  {author} {\bibinfo {author} {\bibfnamefont {N.}~\bibnamefont
  {Aghanim}} \emph {et~al.} (\bibinfo {collaboration} {Planck}),\ }\href
  {\doibase 10.1051/0004-6361/201833910} {\bibfield  {journal} {\bibinfo
  {journal} {Astron. Astrophys.}\ }\textbf {\bibinfo {volume} {641}},\ \bibinfo
  {pages} {A6} (\bibinfo {year} {2020})},\ \bibinfo {note} {[Erratum:
  Astron.Astrophys. 652, C4 (2021)]},\ \Eprint
  {http://arxiv.org/abs/1807.06209} {arXiv:1807.06209 [astro-ph.CO]}
  \BibitemShut {NoStop}%
\end{thebibliography}%

\end{document}